\documentclass[usegraphicx,usenatbib,useAMS]{mn2e}
\newcommand\ion[2]{#1#2} 
\newcommand\lsim{\mathrel{\rlap{\lower4pt\hbox{\hskip1pt$\sim$}}
        \raise1pt\hbox{$<$}}}
\newcommand\gsim{\mathrel{\rlap{\lower4pt\hbox{\hskip1pt$\sim$}}
        \raise1pt\hbox{$>$}}}
\newcommand{\qnamesfourtwo}{J1148+5251}
\newcommand{\qnamestwoeight}{J1030+0524}
\newcommand{\qnamestwotwo}{J1623+3112}
\newcommand{\avenf}{\bar{x}_{\rm HI}^{\rm IGM}}
\newcommand{\fcoll}{f_{\rm{coll}}}
\newcommand{\NHI}{N_{\rm{\ion{H}{I}}}}

\DeclareMathSymbol{\upalpha}{0}{UPM}{"0B}
\DeclareMathSymbol{\upbeta}{0}{UPM}{"0C}

\defcitealias{MH2007}{MH07}

\title[GP Damping Wings in Patchy Reionization]{Evidence of Gunn-Peterson Damping Wings in High--$z$ Quasar Spectra: Strengthening the Case for Incomplete Reionization at $z\sim$6--7}

\author[J. Schroeder, A. Mesinger, and Z. Haiman]
{Joshua Schroeder$^{1}$\thanks{email: jps@astro.columbia.edu; andrei.mesinger@sns.it; zoltan@astro.columbia.edu}, Andrei Mesinger$^{2}$, and Zolt\'an Haiman$^{1}$\\ 
$^1$Department of Astronomy, Columbia University, 550 West 120th Street, New York, NY 10027, USA\\
$^2$Scuola Normale Superiore, Piazza dei Cavalieri 7, 56126 Pisa, Italy}

\begin{document}

\date{}

\voffset-.6in

\pubyear{}

\maketitle

\begin{abstract}
  The spectra of several high--redshift ($z>6$) quasars have shown
  indications for a Gunn-Peterson (GP) damping wing, suggesting a
  substantial mean neutral hydrogen fraction ($\avenf \gsim 0.03$) in
  the $z\approx 6$ intergalactic medium (IGM).  However, previous
  analyses assumed that the IGM was uniformly ionized outside of the
  quasar's \ion{H}{II} region.  Here we relax this assumption and
  model patchy reionization scenarios for a range of IGM and quasar
  parameters.  Compared to uniform reionization, patchy reionization
  imprints a different average damping wing profile with an associated
  sightline-to-sightline scatter.  We quantify the impact of these
  differences on the inferred $\avenf$, by fitting the spectra of
  three quasars: SDSS \qnamesfourtwo\ ($z=6.4189$), \qnamestwoeight\
  ($z=6.308$), and \qnamestwotwo\ ($z=6.247$).  We find that the
  best-fit values of $\avenf$ in the patchy models agree well with the
  uniform case.  More importantly, we confirm that the observed
  spectra favor the presence of a GP damping wing, with peak
  likelihoods decreasing by factors of $\gsim$ few -- 10 when the
  spectra are modeled without a damping wing.  We also find that the
  Ly$\alpha$ absorption spectra, by themselves, cannot distinguish the
  damping wing in a relatively neutral IGM from a damping wing in a
  highly ionized IGM, caused either by an isolated neutral patch, or
  by a damped Ly$\alpha$ absorber (DLA).  However, neutral patches in
  a highly ionized universe ($\avenf\lsim 10^{-2}$), and DLAs with the
  large required column densities ($\NHI \gsim {\rm{ few}}\times
  10^{20}{\rm{ cm}^{-2}}$) are both rare.  As a result, when we
  include reasonable prior probabilities for the line of sight (LOS)
  to intercept either a neutral patch or a DLA at the required
  distance of $\sim 40-60$ comoving Mpc away from the quasar, we find
  strong lower limits on the neutral fraction in the IGM, $\avenf
  \gsim 0.1$ (at 95\% confidence).  This supports earlier claims
  that a substantial global fraction of hydrogen in the $z\approx 6$
  IGM is in neutral form.
 
\end{abstract}


\section{Introduction}

The epoch of reionization, when the radiation from early generations
of astrophysical objects ionized the intergalactic medium (IGM),
offers a wealth of information about cosmological structure formation
and about physical processes in the early Universe.  A variety of
observations in the past few years have provided valuable insight into
this epoch, through measurements of the cosmic microwave background
(CMB) polarization anisotropies, quasar spectra, the kinetic
Sunyaev-Zel'dovich (kSZ) signal and the evolution of the population of
Ly$\alpha$ emitting galaxies.  However, the interpretation of these
observations remains controversial (e.g. \citealt{MH2004,MH2007,
BH2007a, Lidz+07, MFG2009, M2010, MF2008b, McQuinn+08, DMW11, DMF11,
MMS12, Zahn12}).

In particular, fully black Gunn-Peterson (GP) troughs \citep{GP1965},
have been found in a number of $z > 6$ Sloan Digital Sky Survey (SDSS)
quasars, as well as in the quasar ULAS J1120+0641 recently discovered
at $z=7.085$ \citep{Mortlock+11} in the UKIDDS survey \citep{UKIDDS}.
The presence of these troughs suggests that these sources may be
revealing the tail--end of reionization (see, e.g., the review by
\citealt{FCK06}).  However, a neutral fraction as low as $\avenf \sim
10^{-4}$ in a patch of the IGM is sufficient to produce a black GP
trough -- i.e. to make the flux undetectable at the wavelengths
corresponding to resonant Ly$\alpha$ absorption. As a result,
constraints inferred about global properties of reionization -- such
as the volume--averaged neutral hydrogen fraction $\avenf$ -- require
modeling the spatial distribution and time--evolution of neutral
hydrogen in the IGM (see e.g., \citealt{M2010}).

A unique signature of significant neutral hydrogen in the IGM is the
presence of a long tail of absorption, extending to wavelengths far
away from the resonant GP trough -- the so--called GP damping wing
\citep{ME1998}.  The optical depth in the damping wing is $5-6$ orders
of magnitude lower than in the resonant core of the Ly$\alpha$ line.
Therefore this damping wing is detectable only if the neutral fraction
is significant ($\avenf \gsim$ few $\times10^{-2}$).

In the SDSS quasars with a fully black GP trough, flux remains
detectable over a wavelength range of a few $\times 10$ \AA\ blueward
of the Ly$\alpha$ line center, attributable to the nearly
fully-ionized region surrounding the quasars.  This region is highly
ionized, but still contains sufficient residual \ion{H}{I} to produce
observable resonant line-of-sight (LOS) absorption, i.e. the Lyman
$\alpha$ forest. Additionally, as the photons pass through the (possibly) neutral
IGM in the foreground, the absorption in the GP damping wing causes a
further depression of the flux. These two sources of Ly$\alpha$
absorption differ markedly: the resonant absorption within the ionized
bubble strongly fluctuates with wavelength, whereas the absorption
from the GP damping wing smoothly and monotonically increases
blueward, with only a mild wavelength--dependence.  This allows, in
principle, the two different sources of opacity to be separated, given
sufficiently high--quality quasar spectra \citep{CH00,MR00,MHC2004}.
However, any robust interpretation requires modeling due to the
near-degeneracy and uncertainty of the quasar and IGM properties
(e.g., \citealt{MH2004, MH2007, BH2007a, Bhemiss2007, MF2008b,
  MFG2009}).

In the case of the UKIDDS quasar ULAS J1120+0641, the GP trough is
located at a wavelength much closer to the Ly$\alpha$ line center,
i.e. the ionized region appears markedly smaller. As a result, the
situation for this object is somewhat different from the SDSS quasars:
the red GP damping wing can cause significant (10-20\%) absorption on
the red side of the Ly$\alpha$ line \citep{Bolton+11}.\footnote{The
$z=6.44$ quasar CFHQS J0210-0456 \citep{Willott+10b} has a similarly
small ionized region ($\approx$1.7 Mpc) and the same comment applies
to this source; however, this source is intrinsically fainter and has
a lower S/N spectrum.}  Therefore, the GP damping wing and the
intrinsic quasar spectrum must be modeled simultaneously, and care
must be taken to include the uncertainties in the latter (by
comparison, the uncertainty in the intrinsic spectrum plays a
sub-dominant role when deriving lower limits on the neutral fraction
from the SDSS quasar spectra; \citealt{KH09}).  We defer the analysis
of ULAS J1120+0641 to future work.

In our previous analysis of the SDSS quasars (\citealt{MH2007};
hereafter MH07), we modeled the fluctuating resonant absorption within
the ionized zone, detecting signatures of GP damping wing absorption
in two out of three SDSS quasars: \qnamestwoeight\ and \qnamestwotwo.
Treating each as a free parameter, we were able to simultaneously
place limits on the mean neutral fraction $\avenf$, the quasars'
ionizing luminosities, and the sizes of the surrounding ionized
regions.  Most intriguingly, we statistically inferred the presence of
a smooth absorption component corresponding to the GP damping wing,
setting strong lower limits $\avenf \gsim 0.03$ for both
\qnamestwoeight\ and \qnamestwotwo\ (with best--fit values of
$\avenf=1$ in both cases).

In MH07, a uniform ionizing background was assumed.  However, the
topology of reionization is likely very inhomogeneous, with ionized
bubbles growing around highly clustered galaxies
(e.g. \citealt{FZH2004,McQuinn07,TC07, Zahn+2011}).  The shape of the
GP damping wing in a universe with such a ``Swiss--cheese''
reionization topology differs from that in a smoothly ionized IGM (as
could result from more exotic reionization scenarios by X-ray photons;
e.g. \citealt{Haiman11}).

\citet{MF2008b} compared GP damping wing profiles in these two
different scenarios (see also \citealt{McQuinn+08}). Their results
suggest that the difference could be significant, and that it can
introduce a bias and scatter in neutral fraction constraints.  In this
paper, we use physical models for patchy reionization based on a
semi-numerical simulation \citep{MF2007}, and infer constraints on the
neutral fraction and on quasar parameters, using the observed spectra
of the same three quasars that were previously analyzed in MH07.  Our
main goal is to assess whether allowing for patchy reionization
significantly modifies previous results.  Apart from relaxing the
assumption of a uniform ionizing background, our analysis improves on
MH07 by (i) using updated quasar redshift measurements and constraints
on the quasars' ionizing luminosities \citep{CBHB2011}, (ii)
incorporating the fluctuating ionizing background from galaxies near
the quasar, and (iii) a better statistical characterization of the
confidence levels on the model parameters, explicitly contrasting
results from a Bayesian analysis with a parametric bootstrapping
procedure (as opposed to inferring probabilities from a
Kolmogorov-Smirnov test as in MH07).

The rest of this paper is organized as follows.
In \S~\ref{sec:analysis}, we describe our analysis technique,
including a description of the patchy reionization simulations, as
well as a brief summary of how we generate mock absorption spectra and
fit them to the observational data.
In \S~\ref{sec:results}, we present our results.
In \S~\ref{sec:discussion}, we discuss various aspects of our results,
such as the origin of the constraints we obtain, and the robustness of
our results, including the possible presence of damped Ly$\alpha$ systems (DLAs).
In \S~\ref{sec:conclusions}, we summarize our key findings and present
our conclusions.
Throughout this paper, we adopt the background cosmological parameters
($\Omega_\Lambda$, $\Omega_M$, $\Omega_b$, $n$, $\sigma_8$, $H_0$) =
(0.73, 0.27, 0.0455, 1, 0.76, 70 km s$^{-1}$ Mpc$^{-1}$),
consistent with the seven--year results by the {\it WMAP} satellite
\citep{WMAP7}.  Unless stated otherwise, all quantities are quoted in
comoving units.

\newpage

\section{Analysis}
\label{sec:analysis}

Our analysis has four distinct components: (i) running semi--numerical
simulations to model patchy reionization, as well as the spatial
distribution and mass function of halos, and the IGM density and
velocity fields; (ii) using the simulations to create mock absorption
spectra; (iii) modeling the quasars' intrinsic emission spectrum; and
(iv) comparing the observed and simulated absorption statistics.  We
first discuss our patchy reionization models
(\S~\ref{subsec:patchymodels}).  Then in
\S~\ref{subsec:spectrummodels} we present our procedure for creating
mock Ly$\alpha$ absorption spectra (ii).  Finally, we discuss the
comparison of mock and observed spectra (iii-iv) in
\S~\ref{subsec:spectrumfits}.

\subsection{Modeling Patchy Reionization}
\label{subsec:patchymodels}

We use the publicly-available, semi-numerical code
DexM\footnote{http://homepage.sns.it/$\sim$mesinger/Sim.html}
\citep{MF2007} to generate evolved density, halo, and ionization
fields at $z=$ 6.5.  Here we briefly outline the procedure, and refer
the interested reader to \citet{MF2007} for more details.

Our simulation box is $L=250$ Mpc on a side, with a particle grid cell
size of 0.14 Mpc.  Halos are filtered out of the 1800$^3$ linear
density field using excursion-set theory.  Halo locations are then
mapped to Eulerian coordinates at a given redshift using first-order
Lagrangian perturbation theory \citep{Z1970}.  The resulting halo
fields have been shown to match both the mass function and statistical
clustering properties of halos in N-body simulations, well past the
linear regime \citep{MF2007}.

The evolved (non-linear) density field is computed in the same manner,
by perturbing the 1800$^3$ Lagrangian density field.  The resulting
particle locations are then binned onto an Eulerian 600$^3$ grid, and
the corresponding evolved velocity field is re-computed from the
density field.  The final density and ionization fields thus have a
grid cell size of 0.42 Mpc, approximately corresponding to the Jeans
length in the ionized IGM at mean density.  The statistical properties
of the density and velocity fields have been shown to match those from
a hydrodynamic simulation remarkably well \citep{21CMFAST}.
Specifically, the agreement between the density fields at $z\approx7$
is better than $\lsim$10\% for both: (i) the PDF (smoothed on
comparable scales) for roughly a dex around the mean (accounting for
the vast majority of the volume), and (ii) the power spectra at
$k\lsim$ few $\times$ Mpc$^{-1}$.

To generate the ionization field, we use the excursion-set formalism
\citep{FZH2004}, which compares the number of ionizing photons
produced in a region of a given scale to the number of neutral
hydrogen atoms inside that region. Specifically, we identify ionized cells
in our box as those which meet the criterion $\fcoll \geq \zeta^{-1}$,
where $\zeta$ is an ionizing efficiency parameter and $\fcoll$ is the
collapse fraction smoothed around a cell at $(\mathbf{x}, z)$ on
decreasing scales, $R$.  The collapse fraction is computed using the
resolved halo field, including halos down to a minimum mass of $1.7
\times 10^8 M_\odot$, corresponding to the atomic cooling threshold at
$z\sim6$.  The resulting ionization morphologies match those generated
with cosmological radiative transfer algorithms relatively well, with
the power spectra agreeing to within $\sim$ 10--20\% down to the
Nyquist frequency \citep{Zahn+2011}.

We follow the procedure described in \citet{MD2008} to model the
{\it galactic} component of the ionization rate, $\Gamma_{\rm bg}$, inside the
\ion{H}{II} regions, assuming that each galaxies ionizing luminosity is proportional to its halo mass.  At every cell location ($\mathbf{x},z$), we sum the contributions of halos with mass $M_i$ at $\mathbf{x}_i$:
\begin{equation}
\label{eq:flux}
\Gamma_{\rm bg}(\mathbf{x},z) = \frac{(1+z)^2}{4 \pi} \sigma_H \epsilon_{\rm{ion}} \sum_{i} \frac{M_i}{|{\bf x} - {\bf x_i}|^2} ~ e^{-|{\bf x} - {\bf x_i}|/\lambda} ~ ,
\end{equation}
\noindent where $\sigma_H$ is the Lyman limit cross-section, and $\lambda$ is the
ionizing photon mean free path inside HII regions.  For simplicity and computational convenience, our fiducial analysis uses a constant mean free path, $\lambda=60$ (comoving) Mpc, consistent at 1-$\sigma$ with observations at $z\approx6$ \citep{SongailaCowie10}; we return to this point further below.  The rate per unit mass at which ionizing photons are released into the IGM, $\epsilon_{\rm{ion}}$, governs the overall normalization of eq. (\ref{eq:flux}).  This normalization is chosen to match observations (see \citealt{CBHB2011}), resulting in an average ionization rate per hydrogen atom of $\bar{\Gamma}_{\rm bg}=10^{-13}~{\rm s^{-1}}$.  In other words, we fix the {\it mean} galactic ionization rate to match \citet{CBHB2011}, but use individual halo locations to model the local spatial {\it fluctuations}\footnote{For simplicity and to avoid introducing uncertain scalings we use the same UVB for each $\avenf$ -- i.e. we ignore the "shadowing" from the neutral islands in patchy reionization models with $\avenf>0$.  This can become important for large $\avenf$, when the sizes of many ionized bubbles are smaller than the assumed mean free path of $\lambda=60$ Mpc, increasing the spatial fluctuations in the UVB.  However, this is unlikely to influence our results, as the absorption statistics are much more sensitive to the fluctuations in the density field (\citealt{MF2009}; see also \S \ref{subsec:galaxyflux} and the associated discussion).} around this mean, according to eq. (\ref{eq:flux}).

We vary the ionizing efficiency parameter, $\zeta$, to generate a
suite of ionization fields corresponding to various values of $\avenf$
at $z=6$.  Specifically, we use the following eight different values:
$\avenf=\{0.01,0.06,0.10,0.38,0.53,0.64,0.73,0.88\}$.  We then extract
$2\times10^4$ lines of sight (LOSs) centered on the two most massive
halos in the simulation volume ($M\sim3\times10^{12} M_\odot$).  The
LOSs are drawn from randomly chosen directions.  The resulting
density, velocity and ionization LOSs are then used to construct mock
Ly$\alpha$ spectra, as described below.

\subsection{Modeling Quasar Absorption Spectra}
\label{subsec:spectrummodels}

We compute the residual, local neutral fraction along each LOS in the
\ion{H}{II} region surrounding the quasar, whose extent we treat as a
free parameter, $R_S$.  Inside the ionized region, we assume
ionization equilibrium with radiation from the quasar
($\Gamma_{\rm{Q}}$) and the background galaxies ($\Gamma_{\rm{bg}}$),
\begin{equation}
  (\Gamma_{\rm{bg}} +\Gamma_{\rm{Q}}) x_{\rm{\ion{H}{I}}} n_{\rm{H}} = \alpha_B \chi_{\rm eff} n_{\rm{H}}^2 (1-x_{\rm{\ion{H}{I}}})^2,
\label{eq:ionization}
\end{equation}
where $n_{\rm{H}}(\mathbf{x},z)$ is the local (non-linear) number
density of hydrogen atoms, $x_{\rm{\ion{H}{I}}}$ is the local neutral
hydrogen fraction, $\chi_{\rm eff}=1.08$ accounts for additional
electrons from singly-ionized Helium (assuming that the Helium and
Hydrogen ionized fractions are equal with a Helium mass fraction of
$Y_{\rm He}=0.24$), and $\alpha_B=1.7\times10^{-13} \rm{ cm}^3 \rm{
s}^{-1}$ is the case-B recombination coefficient.  The latter is
evaluated at the temperature $T=1.8\times10^{4}$ K (consistent with
recent measurements; \citealt{Bolton2012})\footnote{While case-B
recombination, corresponding to optically thick media, is typically
assumed in the reionization literature, the appropriate coefficient is
somewhat uncertain. Depending on the spectrum of the ionizing sources
and the abundance and density profiles of optically thick Lyman limit
systems, the case-A value could be more appropriate (see discussion in
\citealt{ME2003}).  Instead of attempting to model the appropriate
coefficient, we note that its value is degenerate with both the
assumed temperature, and also with the ionizing luminosity.  In
particular, the difference between the two coefficients is a factor of
$\approx 1.6$ \citep{Osterbrock}, well within the range of possible
temperatures in quasar HII regions (Bolton et al. 2012).}.

Equation \ref{eq:ionization} is solved at every step along the LOS,
using the non-linear density from our simulation.  The contribution to
the ionization rate from the quasar drops off approximately as the
inverse square of the distance, and is given by
\begin{equation}
  \Gamma_{Q}=\frac{1.6\times10^{31}f_\Gamma}{4\pi d_{L}^{2}}
\int_{\nu_{\rm{H}}}^{\infty}d\nu
\left(\frac{\nu}{\nu_{\rm{H}}}\right)^{-1.8}
\left(\frac{1+z}{1+z_{Q}}\right)^{0.8}
\left(\frac{\sigma_\nu}{h\nu}\right)\rm{ s}^{-1}.
\label{eq:gamma_q}
\end{equation}
Here $\sigma_\nu\approx 6.3\times10^{-18}{\rm{ cm}^2}(\nu/\nu_{\rm
  H})^{-3}$ is the frequency-dependent hydrogen ionization
cross-section, $h$ is Planck's constant, $z_{Q}$ is the observed
redshift of the quasar, $z$ is the total redshift corresponding to a
pixel along the LOS, including peculiar velocity [taking
$(1+z)=(1+z_{0})(1+v_{pec}/c)$, where $z_{0}$ is the redshift that
would correspond to the Hubble flow at the comoving distance of the
pixel], $\nu_{\rm{H}}=3.29\times10^{15}$Hz is the ionization threshold
of hydrogen, $f_{\Gamma}$ is the quasar's ionizing luminosity in units
of $1.3\times10^{57}\,\rm{ s}^{-1}$ (following MH07), and
$d_{L}=d_L(z_{0},z_Q)$ is the luminosity distance between the quasar
and the foreground pixel (see, e.g., \citealt{PHW2002}).
Equation~(\ref{eq:gamma_q}) further assumes an intrinsic $\nu^{-1.8}$
power-law spectrum for the ionizing radiation of the quasar, matching
standard quasar templates (e.g. \citealt{TZKD2002}).

As in MH07, our model has three free parameters: (1) the LOS distance
to the edge of the \ion{H}{II} region surrounding the quasar, $R_S$,
(2) the quasar's ionizing luminosity parametrized by $f_{\Gamma}$, and
(3) the mean volume-weighted neutral fraction in the IGM, $\avenf$.
Our model grid covers a different range of $R_S$ for each quasar: in
units of comoving Mpc, for J1148+5251, $R_S = \{37, 39, 41, 43,
45,47\}$, for J1030+0524, $R_S = \{52, 54, 56, 58, 60, 62\}$, and
for J1623+3112, $R_S = \{37, 39, 41, 43, 45,47\}$.  In each case, we
cover the range
$f_{\Gamma}=\{0.1,0.4,0.7,1.0,1.3,1.6,1.9,4.0,6.0,8.0,10.0\}$, and the
eight different values of $\avenf$ listed above.

It is worth pointing out that, in general, for a fixed neutral
fraction, $R_S$ could be determined by the ionizing emissivity and the
lifetime of the quasar, the surrounding galaxies, or a combination of
both.  As such, $R_S$ may be an integral, indirect measure of galaxy
and quasar parameters (e.g., \citealt{WL2007}), but here we simply
take it to directly describe the ionization topology surrounding the
quasar.  We chose a range of $R_S$ values greater than or equal to
the comoving distance between the redshift of the quasar and the
redshift of the first (i.e. reddest) pixel in the quasar's GP-trough.
As we utilize the recently published revised redshift measurements of
\citealt{CBHB2011}, our $R_S$ values are somewhat different (i.e.,
larger) than those in MH07. The wavelengths corresponding to our $R_S$
grid are indicated by the (red) vertical dashed lines in
Figure~\ref{fig:spectra}.

For each given combination of $(R_S, f_{\Gamma},\avenf)$, we create a
mock absorption spectrum corresponding to each LOS extracted from our
simulation box. The spectrum is simply assumed to be black at
wavelengths corresponding to $R > R_S$ and is effectively ignored from
our analysis.\footnote{In principle, spikes of transmission can appear
at $R > R_S$, outside the quasar's zone of influence. The statistics
of the spectrum in this range can also place constraints on the
ionizing background and on the volume-filling fraction of neutral
patches \citep{Croft98, SC02, Barkana02, GCF06}, although in practice,
constraints on the latter require an independent measurement of the
UVB \citep{M2010}.  We emphasize that our results below are
independent of these statistics and arise solely from the GP damping
wing.}  At $R < R_S$, the total optical depth includes resonant
absorption from neutral HI inside the ionized region, as well as
absorption through the damping wing by neutral hydrogen located
outside $R>R_S$.  We first create an array of observed wavelength bins
$\lambda_j$, and for each bin, we sum the optical depth from every
pixel $i$ along the LOS,
\begin{equation}
\tau_{j}={\sum_{i}}\frac{\sqrt{\pi}e^{2}f_{\alpha}}{m_{e}c}x_{i}n_{i}\frac{\delta s}{1+z_{0,i}}\phi_{\alpha}\left(\frac{\lambda_{j}}{1+z_i}\right),
\label{intlineoptdepth}
\end{equation}
where $e=4.8\times10^{-10}$ esu is the charge of the electron,
$f_{\alpha}=0.4162$ is the Ly$\alpha$ oscillator strength,
$m_{e}=9.1\times10^{-28}$ g is the mass of the electron, 
$z_i$ is the total redshift of pixel $i$, including its peculiar
velocity, $x_i$ is the neutral fraction and $n_i$ is the total
hydrogen density in pixel $i$, $\delta s/(1+z_{0,i})$ is the
(physical, as opposed to comoving) width of the pixel, and
$\phi_{\alpha}$ is the normalized absorption profile.

The frequency-dependence of the Ly$\alpha$ line is
\citep[][eq. 23.97]{Peebles1993}
\begin{equation}
\phi\left(\omega\right)=
\frac{\Lambda_\alpha\left(\omega/\omega_{\alpha}\right)^{4}}{\left(\omega-\omega_{\alpha}\right)^{2}+\Lambda_\alpha^{2}\left(\omega/\omega_{\alpha}\right)^{6}/4},\label{eq:Peeblesprofile}
\end{equation}
where $\Lambda_\alpha=6.25\times10^8{\rm s^{-1}}$ is the radiative
decay rate.  A common approach in the literature is to use the
convolution of a Lorentzian \citep[][eq. 23.43]{Peebles1993} with the
thermal broadening, i.e. a Voigt profile \citep{Press1993} near the
resonance $\omega\approx\omega_{\alpha}$ and an analytic solution for
the damping wing far away from resonance that ignores Doppler
broadening but includes the $\left(\omega/\omega_{\alpha}\right)^{4}$
Rayleigh-scattering tail \citep{ME1998}.  As our results may depend on
intermediate regimes between the damping wing and line center, in
principle, the thermally broadened line profile should be computed by
convolving equation~(\ref{eq:Peeblesprofile}) with the
temperature-dependent Doppler profile. In practice, we have
approximated this full convolution with
\begin{equation}
\phi_\alpha\approx\phi_{\rm voigt}\left(\omega,T\right)\times
\left(\omega/\omega_{\alpha}\right)^{4},\label{eq:Peeblesdopplerapprox}
\end{equation}
where $\phi_{\rm voigt}$ is the Voigt profile for our temperature $T$.
This approximation approaches that given by \citet{ME1998} on the far
red side of the damping wing.

In our patchy reionization models, specifying $\avenf$ defines the
ionization morphology.  It is unlikely that the edge of the
surrounding HII region will happen to lie at any given fixed distance,
$R_S$.  Therefore, in order to keep $R_S$ as an independent free
parameter, we ``shift'' the ionization array (keeping the density
array along the LOS fixed) by the minimal amount so that a neutral
pixel falls at the location at $R_S$.  However, in the highly-ionized
models, we often encounter LOSs without any neutral pixels.  In these
cases, we keep the density and velocity field along the LOS, but we
``recycle'' the ionization array from another, randomly chosen LOS to
make sure that a neutral pixel is located at $R_S$.  This shifting is
done in order to treat $R_S$ as a free parameter, while preserving the
ionization morphology from the simulation at $R>R_S$ (which determines
the GP damping wing).  After presenting these conservative
constraints, we then combine them with the (model-dependent)
probability that a neutral pixel is, in fact, found at the radius
$R_S$ (see the discussion in \S \ref{sec:res_patchy} below).  This
takes into account that a more ionized universe generally has larger
HII regions by introducing a prior on $R_S$.

\subsection{Model Parameters}
\label{sec:params}

In summary, our fiducial analysis has three free parameters (analogous to MH07):
\begin{itemize}
\item {\bf $\avenf$} --  the mean volume-weighted neutral fraction in the IGM. At a given redshift, $\avenf$ is a function of $\zeta$ and $T_{\rm vir}$ (see below)\footnote{In principle, the ionizing morphology at fixed $\avenf$ can also be a function of $\lambda$; however, this dependence is very weak for $\lambda\gsim10$ Mpc (e.g. \citealt{McQuinn07}).}.\\
\item {\bf $f_\Gamma$} -- the quasar's ionizing luminosity in units of $1.3\times10^{57}\,\rm{ s}^{-1}$. \\
\item {\bf $R_S$} -- the LOS distance to the edge of the \ion{H}{II} region surrounding the quasar.
\end{itemize}
\noindent Our models include additional parameters, which we either do not vary or which are used to derive the above quantities.  Ideally one would like to explore as large a parameter space as possible; however, in practice, computational costs have limited us to vary at most three parameters simultaneously.   In Sections 4.4--4.6 below, we perform additional investigations to assess the robustness of our results to degeneracies missing in our three--parameter models.   The additional parameters include:
\begin{itemize}
\item {\bf $\zeta$} -- the ionizing efficiency of high-redshift galaxies.  This quantity can be defined as $\zeta = f_{\rm esc} f_\ast N_\gamma / (1+n_{\rm rec})$, where $f_{\rm esc}$ is the fraction of ionizing photons produced by stars that escape into the intergalactic medium (IGM), $f_\ast$ is the star formation efficiency, $N_\gamma$ is the number of ionizing photons per stellar baryon, and $n_{\rm rec}$ is the mean number of recombinations per baryon.  For reference, $f_{\rm esc}=0.1$, $f_\ast=0.1$, $N_\gamma =4000$ (appropriate for PopII stars), and $n_{\rm rec}=1$ yield $\zeta = 20$.  As described above, we vary $\zeta$ to generate ionization maps, computing the resulting $\avenf$.\\
\item {\bf $T_{\rm vir}$} --  the minimum virial temperature of halos hosting star-forming galaxies.  As is common in reionization literature, we set this to the atomic-cooling threshold, $T_{\rm vir}=10^4$ K.  Although $T_{\rm vir}$ could have a somewhat different value, the ionization morphology at fixed $\avenf$ (the quantity of interest in this work) is fairly robust to (reasonable, physically-motivated) changes in $T_{\rm vir}$ \citep{McQuinn07, MMS12}.\\
\item {\bf $\bar{\Gamma}_{\rm bg}$} -- the mean value of the galactic UVB.  As mentioned above, we set this value to $\bar{\Gamma}_{\rm bg}=10^{-13}$ s$^{-1}$.  Although this matches observations of the $z\sim6$ Lyman alpha forest (e.g. \citealt{CBHB2011}), such measurements have large uncertainties.   Luckily, our conclusions are not sensitive to the UVB, as we discuss below.\\
\item {\bf $\lambda$} -- the ionizing photon mean free path inside HII regions, generally set by Lyman limit systems (LLSs).  We use a fiducial value of $\lambda=60$ (comoving) Mpc, consistent at 1-$\sigma$ with observations at $z\approx6$ \citep{SongailaCowie10}.  The mean free path is used to compute the fluctuations in $\Gamma_{\rm bg}$, the galactic UVB\footnote{It is also possible that LLSs along the line of sight from the QSO attenuate the quasar's flux, $\Gamma_Q$ so that it deviates from the canonical $\propto r^{-2}$ behaviour.  Such LLSs would need to have column densities low enough to escape detection in HIRES and ESI.  We return to this issue in \S \ref{sec:rsquared}.}.
  In principle, $\lambda$ is poorly constrained at high redshifts, and could even vary spatially (e.g. \citealt{Crociani11}).  However, our conclusions are not very sensitive to $\Gamma_{\rm{bg}}$, for several reasons.  First, for most of the relevant spectral range and parameter choices, the quasar's flux dominates the ionizing background, with galaxies becoming important only in the relatively unimportant high-$R_S$, low-$f_\Gamma$, low-$\avenf$ regime (see more discussion of this point below in \S~\ref{subsec:galaxyflux}, and also \citealt{Lidz2007} and \citealt{BH2007a}).  Second, the {\it profile} of $\Gamma_{\rm{bg}}$ (as a function of the distance from the quasar) is not degenerate with any of our three free parameters\footnote{$\Gamma_{\rm bg}$ is spatially fluctuating around the mean value, with an expected variance on the scale of our cell size of a factor of $\sim$few \citep{Lidz+07, MD2008, MF09}.  There is a slight evolution of the mean with distance from the quasar, due to the biased nature of the clustered galaxies (which is accounted for in our models).  The fact that none of our free parameters have similar imprints on the QSO spectra suggests that any inaccuracies in our modeling of $\Gamma_{\rm bg}$ are unlikely to have a strong bias on our results.}.  Finally, as mentioned above, the average transmission is far more sensitive to the density field than to inhomogeneities in the background flux \citep{MF2009}.
\end{itemize}

\subsection{Fitting the free parameters in our model}
\label{subsec:spectrumfits}

\subsubsection{Pixel Optical Depth Distributions}
\label{subsubsec:tauPDFs}

\begin{figure}
\includegraphics[scale=0.5]{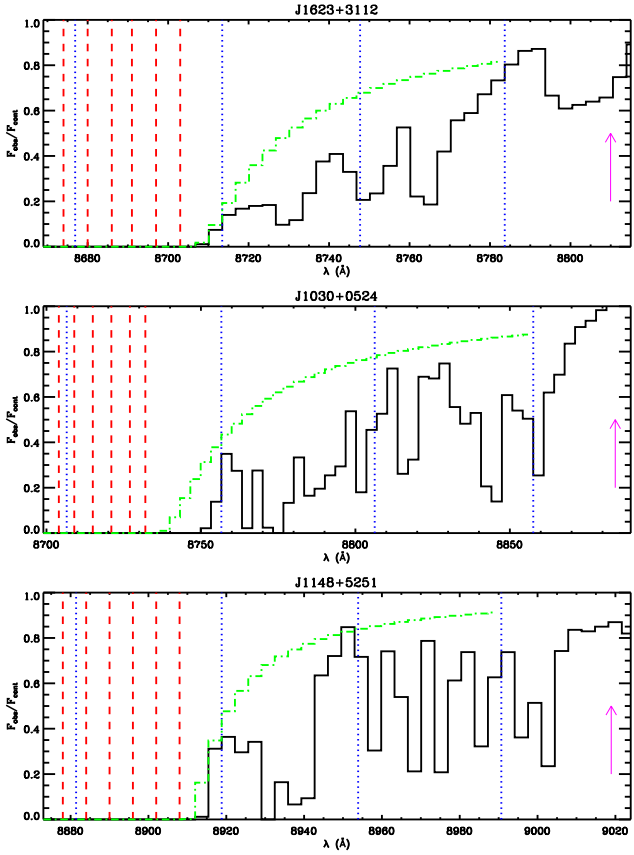}
\caption{\label{fig:spectra} The continuum--normalized spectra of the
  three SDSS quasars used in this paper. The dotted (blue) vertical
  lines in each panel mark the boundaries of the three wide
  wavelengths bins used in the analysis, while the dashed (red) lines
  show the wavelengths corresponding to various assumed sizes ($R_S$)
  of the quasar's ionized region explored in our models.  The
  dot-dashed (green) curves indicate, for illustrative purposes only,
  the transmission attributed to the GP damping wing alone. These are
  the {\em mean} damping wings in the best-fit (without an R$_S$ prior) models for each quasar,
  but our actual analysis includes the LOS-to-LOS scatter in both the
  damping wing and the resonant absorption).  The magenta arrow on the
  right of each panel marks the location of the Ly$\alpha$ line at
  each quasar's redshift.}
\end{figure}

In order to compare our simulated spectra to the observations, we also
need an intrinsic (unabsorbed) quasar emission template. These are
obtained by fitting the Keck ESI observations of the quasars (e.g.,
\citealt{WBFS2003}) only on the red side of the Ly$\alpha$ emission
line, where IGM absorption is minimal.  The intrinsic spectral fits
are taken from MH07, to which we refer the interested reader for
details.  Briefly, the model fit to the data consists of the sum of a
power-law continuum, a double-Gaussian Ly$\alpha$ emission line, and a
\ion{N}{V} emission line.  The observed continuum-normalized flux in
wavelength bin $j$ is then obtained as
\begin{equation}
\widetilde{F}_{obs,j}=F_{obs,j}{/{F_{cont,j}}}\label{Fobs} ~ ,
\end{equation}
where $F_{cont,j}$ and $F_{obs,j}$, are the intrinsic and observed
flux, respectively.  We compare this observational signal with our
mock continuum-normalized spectra, given by
\begin{equation}
\widetilde{F}_{sim,j}=\exp(-\tau_{j})\label{Fsim}.
\end{equation}
Simulated optical depths can be far in excess of what is observable,
so to ensure a fair comparison between simulation and observation, we
impose a floor of minimum flux given by the observational errors
($\sigma_{obs,j}$, taken from \citealt{WBFS2003}):
\begin{equation}
\widetilde{F}_{min,j}={A\sigma}_{obs,j}{/F}_{cont,j}\label{Fmin},
\end{equation}
with the constant chosen to be $A=3$ for our analysis, corresponding
to 3-$\sigma$ detections.  For all values of
$\widetilde{F}_{j}<\widetilde{F}_{min,j}$ we treat the simulated and
observed continuum-normalized flux to be the same as zero flux. This
floor corresponds to a maximum observable optical depth ranging
between $\tau_{obs}=4.75$ to 5.5. The continuum-normalized spectra of
the three quasars, subject to these conditions, are reproduced in
Figure~\ref{fig:spectra}.  Since present-day simulations do not have
the dynamic range to model biases surrounding the bright, rare
quasars, while at the same time resolving the Lyman-$\alpha$ forest,
we excise the region 25 \AA\ blue-ward of the Ly$\alpha$ peak from our
analysis.  This roughly corresponds to the expected mean radius ($\sim
10$ comoving Mpc) of the large--scale overdensity surrounding such
quasars (e.g. \citealt{BLGalaxyFormation2004}).

We then compare the observed spectra to simulated mock spectra, using
the two-sided Kolmogorov-Smirnov (KS) distance statistic $D_{KS}$
\citep{NumRec1992}, applied to the observed vs. simulated flux density
PDFs.  Each spectrum was binned into three wide wavelength ranges,
with two of the bins chosen to lie red-ward of the pixels where the
observed spectrum is consistent with zero flux -- nominally
corresponding to the apparent edge of the ionized bubble surrounding
the quasar. The apparent edge falls within the third, bluest bin (see
Figure~\ref{fig:spectra}).  Within each of these three bins, the PDFs
of the optical depth was obtained, for a given combination of
$(R_S,f_{\Gamma},\avenf)$, using all LOSs and all pixels within the
bin.  Each of the three model PDFs were then compared to the
corresponding optical depth PDFs inferred from the observed spectra.

\subsubsection{Best-Fit Models and Confidence Limits}
\label{subsubsec:parameterfitting}

In MH07, for each set of model parameters, the product of the three KS
probabilities associated with the three $D_{KS}$ values, was used as
the likelihood of that model.  The probability $Q_{KS}(D_{KS})$ is
given by the commonly used approximate analytic fitting formula
\citep{NumRec1992}.  However, this approach is potentially
problematic, since: (i) the analytic formula is only accurate for
``well-behaved'' distributions; (ii) $\tau$ values in nearby pixels
can be correlated, and (iii) the $\tau$ values in each of the $\sim$10
consecutive pixels within a bin are drawn from slightly different
underlying probability distributions.

To address the impact of all of these issues, we explicitly compute
the PDF of the KS distance statistic $D_{KS}$ in each of our
models. We first compute the pixel optical depth PDF for the entire
set of $2 \times 10^4$ LOSs - this constitutes our best estimate for
the $\tau$ PDF predicted in a given model.  We then compare this
prediction to the $\tau$ PDF for each individual LOSs in the same
model, thus obtaining the (cumulative) PDF of $D_{KS}$ in this model.
We have found that these CPDFs differ significantly from the fitting
formula $Q_{KS}$ in \citet{NumRec1992}. We traced these differences to
the two reasons (i) and (ii) above. First, the floor imposed by
equation~(\ref{Fmin}) causes a pile-up of nearly identical optical
depths (corresponding to the number of ``black'' pixels) in the $\tau$
PDF.  This produces sharp steps in the sample CPDFs, and this unusual
feature modifies the PDF of $D_{KS}$ (this effect is similar to a
decrease in the number of effective data points).  Second,
correlations between neighboring pixels skew the CPDF away from
$Q_{KS}$.  We demonstrated this by first masking out the black pixels
from the analysis, and then picking $\tau$ values from random LOSs to
generate $2 \times 10^4$ uncorrelated $\tau$ distributions. In this
case, the CPDF of $D_{KS}$ indeed was consistent with
$Q_{KS}(D_{KS})$.

For each $(R_S,f_{\Gamma},\avenf)$ model (and in each of the three
spectral bins $i$ of each quasar) we measure the KS distance
$\bar{D}_{KS,obs}$, by comparing the $\tau$ CPDFs in the given model
(averaged over all LOSs) with the corresponding $\tau$ CPDF in the
quasar spectrum.  We then construct the expected CPDF,
$P_{(R_S,f_{\Gamma},\avenf)}^{{\rm bin}~i}(>D_{KS})$, of the KS
distance in this model, by comparing the $\tau$ CPDF of each LOS to
the mean $\tau$ CPDF constructed from all LOSs. The probability that
$D_{KS}$ exceeds $\bar{D}_{KS,obs}$,
$P_{(R_S,f_{\Gamma},\avenf)}^{{\rm bin}~i}(>\bar{D}_{KS,obs})$, is
assigned to this model.  Finally, multiplying the three probabilities
in the three wavelength bins gives the model's overall likelihood
$p(R_S,f_{\Gamma},\avenf) = \prod_{i=1}^{3}
P_{(R_S,f_{\Gamma},\avenf)}^{{\rm bin}~i}(>\bar{D}_{KS,obs})$ of the
model. The best-fit model was identified as the one that maximizes
this overall likelihood.

We next place confidence limits around this best-fit model by
interpreting $p(R_S,f_{\Gamma},\avenf)$ directly as Bayesian
probability {\em densities}, and integrating these probability
densities over our parameter space to obtain the confidence intervals.
The results from this approach depend on the (arbitrary)
parameterization of the parameter space volume. Here we chose
$R_S,\log f_{\Gamma}$ and $\log \avenf$ -- effectively adopting a
uniform prior on these quantities.  As a check on the robustness of
these Bayesian confidence limits, we compare these results to those
inferred from a parametric bootstrapping procedure (see
\S~\ref{subsec:bootstrapping}).

\section{Results}
\label{sec:results}

\subsection{Uniform Reionization}
\label{sec:res_uniform}

Before attempting to improve on the results in MH07, we first
reproduced their main results, assuming, as in that paper, that the
IGM outside $R_S$ is ionized by a spatially uniform background.  A few
of the details of our analysis, however, still differ from MH07:

(1) To accommodate the revisions in the redshifts of the three quasars
(by up to $\Delta z\approx 0.03$; \citealt{Carilli+10}), we modify the
boundaries of our three wavelength bins compared to those used by
MH07.  Likewise, we slightly adjust and expand the set of values for
$R_S$ used in our grid of models.  In particular, for J1030+0524,
which has a Ly$\beta$ trough that lies at a significantly larger
distance from the quasar than the Ly$\alpha$ trough, we impose a
minimum size for the ionized region to $R > 52$ Mpc, corresponding to
the location of the Ly$\beta$ trough (as opposed to 40 Mpc in MH07,
using the old redshift).

(2) Rather than having a uniform UVB inside the quasar's HII region
($R<R_S$), we compute the inhomogeneous UVB from the nearby galaxies.

(3) We find the best-fit models using our own determination of the KS
probabilities, rather than the standard fitting formula, as discussed
in \S~\ref{subsubsec:parameterfitting} above.

(4) We add a strong prior for the probability distribution of the
quasar's ionizing flux $f_{\Gamma}$, based on the best--fit values and
the errors obtained by \citet{CBHB2011}.  In practice,
\citet{CBHB2011} quote errors on the logarithmic quasar luminosity
$\ln L_Q$. For simplicity, we multiply the overall likelihood
$p(R_S,f_{\Gamma},\avenf)$, as defined in the previous section, by a
Gaussian in $\ln f_{\Gamma}$, centered at $\ln
f_{\Gamma}=(0.982,0.536,1.58)$ for \qnamestwoeight, \qnamestwotwo\ and
\qnamesfourtwo, respectively, and with the same uncertainty of
$\sigma_{\ln f}=0.31$ for all three quasars.

(5) MH07 quoted the values of the parameters where the KS
probabilities fell to $1/3$rd, $1/9$th, and $1/27$th of their peak
values.  For a 1D Gaussian, these would correspond to $\approx 1.5$,
$\approx 2$, and $\approx 2.5\sigma$ confidence limits. The procedures
discussed above allow us to integrate the likelihoods within the fixed
contours, and to compute actual confidence limits in three
dimensions. This is a particularly important change: due to the long
non-Gaussian tails, and having a three-dimensional parameter space, we
find that the confidence levels differ significantly from the simple
1D Gaussian expectation.

In Table~\ref{table1}, we show the parameter combinations where the
likelihoods peak, as well as the best-fit models returned by the
bootstrapping procedure. The first row for each quasar lists the
results for the uniform-ionization case. For comparison, the values
from MH07 are listed in the second row.  The likelihoods peak at high
neutral fractions for all three quasars, though none precisely at
$\avenf = 1.0$. A notable difference from MH07 can be seen in the case
of J1148+5251: the KS probability peaks at a higher neutral fraction
($\avenf=0.88$ vs. 0.16), though this value is still consistent with
the $1/3$rd probability contour of MH07. We also find significantly
higher $R_S$ values for both J1030+0524 and J1623+3112 compared to
MH07, caused by the increase in the intrinsic redshifts of these two
quasars. This also leads to a factor of $\sim 2-3$ larger inferred
quasar luminosities, which is not surprising: the material at the same
observed wavelength is now farther from the quasar, so a larger
intrinsic luminosity is needed to produce similar absorption.

When we use the same $Q_{KS}$ statistic as MH07, we further recover
very similar lower limits on $\avenf$, attributable to GP-damping
wings for both J1030+0524 and J1623+3112. As in MH07, we also find
that low values of $\avenf$ can not be excluded for J1148+5251.

We then switch to the new KS probability densities and integrate these
over the parameters to obtain confidence levels.  Our 68\% and 95\%
two-dimensional confidence contours for each of the three quasars are
shown in Figure~\ref{fig:uniform}.  The constraints shown in each
panel have been marginalized over the third parameter.  Focusing on
the further marginalized constraint on the single parameter $\avenf$,
we obtain a clear lower limit for \qnamesfourtwo\ and \qnamestwotwo\
at about $\avenf \gsim 0.04$ while \qnamestwoeight\ has a somewhat
stronger lower-limit at $\avenf \gsim 0.07$ .

\subsection{Patchy Reionization}
\label{sec:res_patchy}

We next fit the spectra using our patchy reionization models.  Unless
stated otherwise, all constraints quoted below correspond to 95\%
confidence limits.

\subsubsection{Results with a Uniform Prior on $R_S$}
\label{sec:nopriors}

In Figure~\ref{fig:patchy}, we present the joint 2D 68\% and 95\%
confidence contours for each quasar.  As in the uniform reionization
models, the best fits are located at high neutral fractions ($\avenf
\ge 0.38$).  The most conspicuous change, however, is that the
contours expand to significantly lower values of $\avenf$, with the
lowest value $\avenf=0.01$ still falling inside the 95\% CL on the
$\avenf-f_\Gamma$ plane.  The single-parameter marginalized constraint
yields essentially the same lower limit, weakened to $\avenf \gsim
0.02$ for all three quasars.

\subsubsection{Adding a Physical $R_S$ Prior}
\label{sec:priors}

As described in \S~\ref{subsec:patchymodels}, we shifted the
ionization field in the patchy--reionization simulation, in order to
ensure that the quasar's ionization front can encounter a neutral
pixel at the pre-specified location $R_S$.  This was necessary to
build up the statistics of the pixel optical depth PDF, and to be able
to treat $R_S$ as one of our free parameters.  However, as described
above, these shifts make fits to the observed quasar spectra unfair:
clearly, it is less likely to find a neutral pixel within a fixed
radius $R_S$ in a highly ionized universe than in a more neutral
universe.  Thus, we next add in our analysis the prior probability for
the LOS to intersect a neutral patch within $R_S$, taken from the
patchy reionization simulations.  These probabilities are shown in
Figure ~\ref{fig:Rsprior}.

In Figure~\ref{fig:patchyprior} we present the confidence regions with
these prior probabilities included.  Unsurprisingly, adding the priors
has the effect of excluding models with lower neutral fractions. For
each of the three quasars, the 95\% contour excludes models with
$\avenf \lsim 0.05$.  The single-parameter marginalized constraint
yields $\sim 7$--9 times tighter limits, with \qnamesfourtwo\ and
\qnamestwotwo\ at $\avenf\gsim 0.14$ and \qnamestwoeight at
$\avenf\gsim 0.11$.  We consider this to be the ``fairest'' statistic,
and present these constraints as the main result of this work.

\section{Discussion}
\label{sec:discussion}

\subsection{The Detection of the GP Damping Wing}
\label{subsec:dampingwings}

The main result of this paper is the strong lower limit on the mean
neutral hydrogen fraction, $\avenf \gsim 0.1$.  The mean GP damping
wing absorption profiles, averaged over all LOSs in our best-fit
models (in the patchy reionization case, with the $R_S$ prior
included, and with the best-fits corresponding to the peak of the
$P(D_{KS})$ likelihood) are shown, for illustrative purposes, by the
dot-dashed (green) curves in
Figure~\ref{fig:spectra}.\footnote{Interestingly, the damping wings
produce some absorption, at the 5-10\% level, extending to the red
side of the quasar's Ly$\alpha$ emission line.  This absorption was
neglected in our modeling of the quasar's intrinsic spectrum.  Taking
this absorption into account would increase the intrinsic flux; since
an overall increase is degenerate with $f_{\Gamma}$, we do not expect
this to change our conclusions on the inferred damping wing.}  We wish
to confirm that our constraints arise from the statistical preference
for these GP damping wings in the observed spectra, and not because of
some other feature.\footnote{For example, had we not blacked out our
model spectra in the region $R>R_S$, we would have been sensitive to
the absorption statistics in the IGM, as mentioned in a footnote in
\S~\ref{subsec:spectrummodels}.}  To test this explicitly, we first
re-fit the spectra, but this time ignoring the absorption by the GP
damping wing.  At the location of the best-fit model (i.e. with $R_S$
and $f_\Gamma$ fixed at the values listed in Table~\ref{table1}), the
KS probabilities decrease by more than an order of magnitude for
\qnamestwoeight\ and \qnamestwotwo\ and more than two orders of
magnitude for \qnamesfourtwo.  This demonstrates the sensitivity of
the data to the presence/absence of the damping wings.

Even if the data is sensitive to the damping wings, however, our fits
might not be sensitive to their presence, owing to degeneracies with
other parameters. Indeed, we found that when we allow $R_S$ and
$f_\Gamma$ to float, the best-fit value of $f_\Gamma$ was lower for
\qnamestwotwo\ and \qnamesfourtwo\ than in our fiducial models which
include a damping wing. This is unsurprising -- when the quasar flux
is lower, the resonant opacity is increased and can compensate for the
loss of the damping-wing absorption.  However, in these
damping-wing-less best-fit models, the peak probabilities are still
factors of $\sim$6--11 below those in the original best-fit
models. This demonstrates that our other parameters, e.g. a lower
quasar ionizing flux, cannot fully mimic the presence of a GP damping
wing.

\subsection{Parametric Bootstrapped Confidence Contours and Scatter in the Damping Wing}
\label{subsec:bootstrapping}

The confidence contours shown in Figures~\ref{fig:uniform},
~\ref{fig:patchy}, and \ref{fig:patchyprior} rely on the assumption of
an underlying flat prior on $\log\avenf$ (as well as either a flat or
a physical prior on $R_S$, and a Gaussian on $\log f_\Gamma$).  Since
the choice of the flat prior on $\log\avenf$ is essentially arbitrary,
we have replicated our entire analysis with an alternate, parametric
bootstrapping procedure.  Namely, we repeat the analysis discussed in
\S~\ref{subsubsec:parameterfitting} another $2\times10^4$ times,
replacing the observed spectrum with the mock spectra generated along
each of the $2\times10^4$ LOSs in our best-fit model.  This results in
a set of $2\times10^4$ new best--fit parameter combinations, which, in
general, differ from the input values.  The number of the new best-fit
values in our 3D parameter space then directly yields the joint
confidence levels.  This method does not require an explicit prior on
how the parameters are distributed, and is a common technique to
estimate confidence levels \citep{NumRec1992}.

The results of this procedure are shown in
Figure~\ref{fig:patchypriorboot}.  The analysis includes the physical
$R_S$ prior from Figure~\ref{fig:Rsprior} (here used only to identify
the initial best-fit models from which the LOSs are bootstrapped), and
so this figure is to be compared to Figure~\ref{fig:patchyprior}.  As
this comparison reveals, the bootstrapping returns noticeably tighter
contours.  The single-parameter constraints on $\avenf$ have tightened
to $\avenf\gsim 0.48$ for \qnamestwoeight\ and $\avenf \gsim 0.62$ for
\qnamestwotwo\ and \qnamesfourtwo. These constraints are weaker
without the $R_S$ prior, in which case the lower limit for
\qnamestwoeight\ is $\avenf\gsim 0.11$, \qnamestwotwo\ is $\avenf\gsim
0.07$, and \qnamesfourtwo\ is $\avenf\gsim 0.14$.  The lower limits
derived for all three quasars in the corresponding Bayesian approaches
are much weaker.

The large differences between the bootstrapped and the Bayesian
confidence regions has a simple physical interpretation.  The
bootstrapping procedure samples the LOSs in our original best-fit
models, and therefore the extent of the confidence regions reflect the
LOS-to-LOS variations of the mock spectra in these best-fit models.
For each quasar, this best-fit model corresponds to a nearly neutral
IGM.  Most importantly, in these nearly-neutral models, there is
relatively little scatter in the strength of the damping wings --
every LOS in each model contains a wide swath of neutral IGM.  In
contrast, low-$\avenf$ models contain only relatively small and
isolated neutral islands. For example, in the model with
$\avenf=0.01$, we find that neutral patches have a median diameter of
only $\sim 7.5$ Mpc, and are typically separated by distances of order
$\sim 100$ Mpc (see also \citealt{M2010}). It is therefore very
unlikely to find any LOS in our high-$\avenf$ models, that is best fit
by the narrow and weak damping wings in low-$\avenf$ models.  This
results in the tight constraints generated by the bootstrapping
procedure.

The situation is quite different when we estimate confidence levels
directly from the Bayesian likelihoods.  In this case, the probability
density at each point in the parameter space is estimated using the
LOS-to-LOS variations {\em in that model}, which generate the $\tau$
CPDFs.  As a result, low-$\avenf$ models are ruled out at a much
weaker significance: these models have a large scatter.  For example,
in the model with $\avenf=0.01$, we find that $\sim 5\%$ of the
sightlines contain much larger neutral patches, with a diameter $\gsim
35$ Mpc. We conclude, therefore, that the low-$\avenf$ models have a
relatively wide probability distribution of the KS-distance $D_{KS}$,
and are therefore difficult to rule out at a high significance.

In principle, the scatter in resonant absorption may be important as
well: the sightlines in the low-$\avenf$ models that look more similar
to the QSO (i.e. with $D_{KS}<\bar{D}_{KS,obs}$) might be those that
have unusually high neutral density inside the HII region. To study
this, we compare a sub-sample of ``good-fit'' LOSs in our
$\avenf=0.01$ model (i.e. LOSs having low values of
$D_{KS}<\bar{D}_{KS,obs}$) to the full sample of $2\times10^4$
sightlines. The low--$D_{KS}$ sub-sample has a noticeably stronger
damping wing than the full sample.  In contrast, the median resonant
optical depths between the two samples are indistinguishable.  We
therefore confirm that the damping wing statistics are driving our
conclusions and that the scatter in the damping wings make the
Bayesian and the bootstrapping confidence limits different.  These
conclusions also show that the {\em scatter} in the damping wing
strengths and shapes has a strong dependence on the average global
ionization $\avenf$, and imply that this can provide an additional
probe of reionization \citep{MF2008b}.

Apart from the three quasars analyzed here (as well as the $z=7.1$
source ULAS J1120+0641) with full GP troughs, currently there are
$\sim 20$ quasars known at $z\gsim 6$ \citep{Willott+10, Carilli+10}.
The sample is quite heterogeneous, and many quasars are too faint for
useful spectroscopy. However, there are a handful of sources with
spectra whose quality is similar to the quasars we analyzed, and yet
they do not show a fully black GP trough. This may place an upper
limit on $\avenf$, since open sightlines, with no neutral patches,
become rare in a nearly neutral universe.

\subsection{Parameter Dependencies and Degeneracies}
\label{subsec:parameters}

\begin{figure}
\begin{tabular}{c}
\includegraphics[scale=0.75]{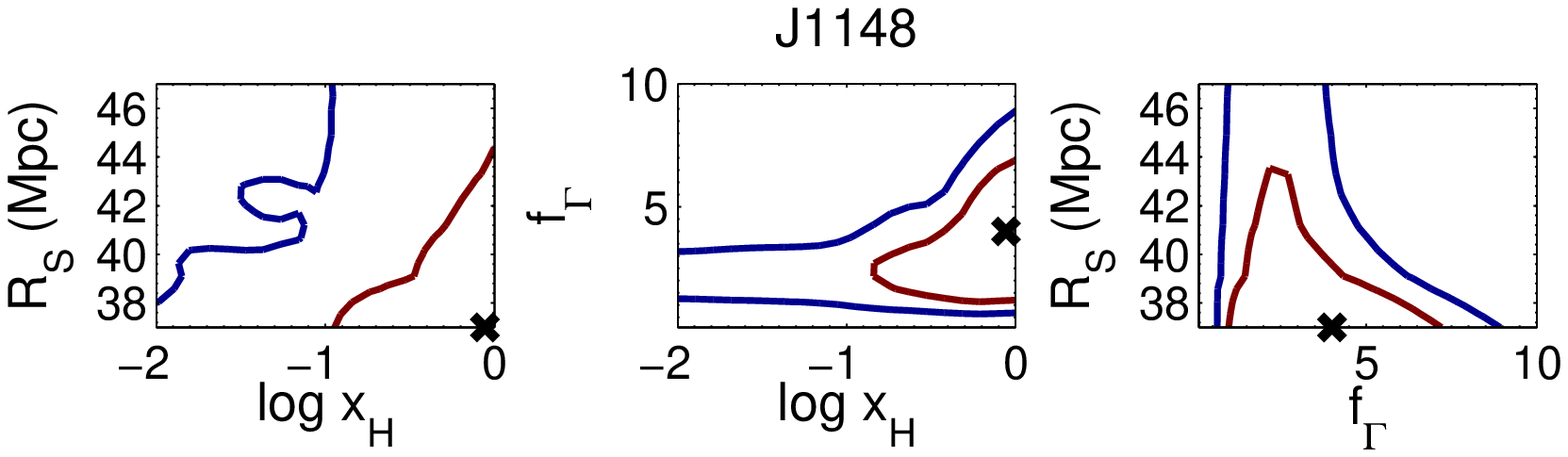}\tabularnewline
\includegraphics[scale=0.75]{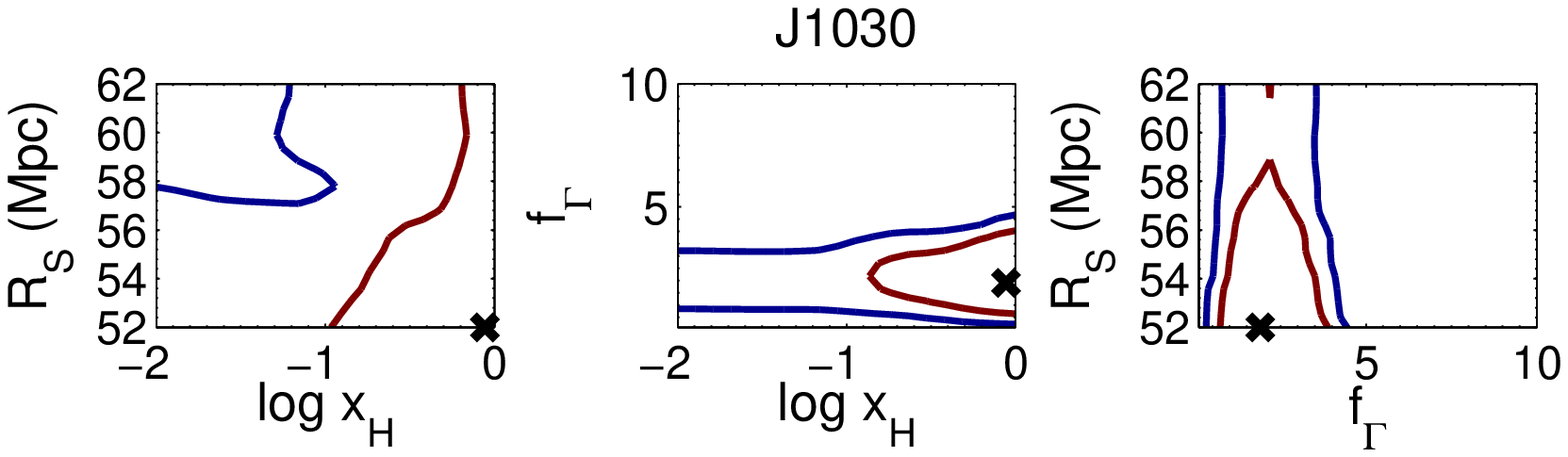}\tabularnewline
\includegraphics[scale=0.75]{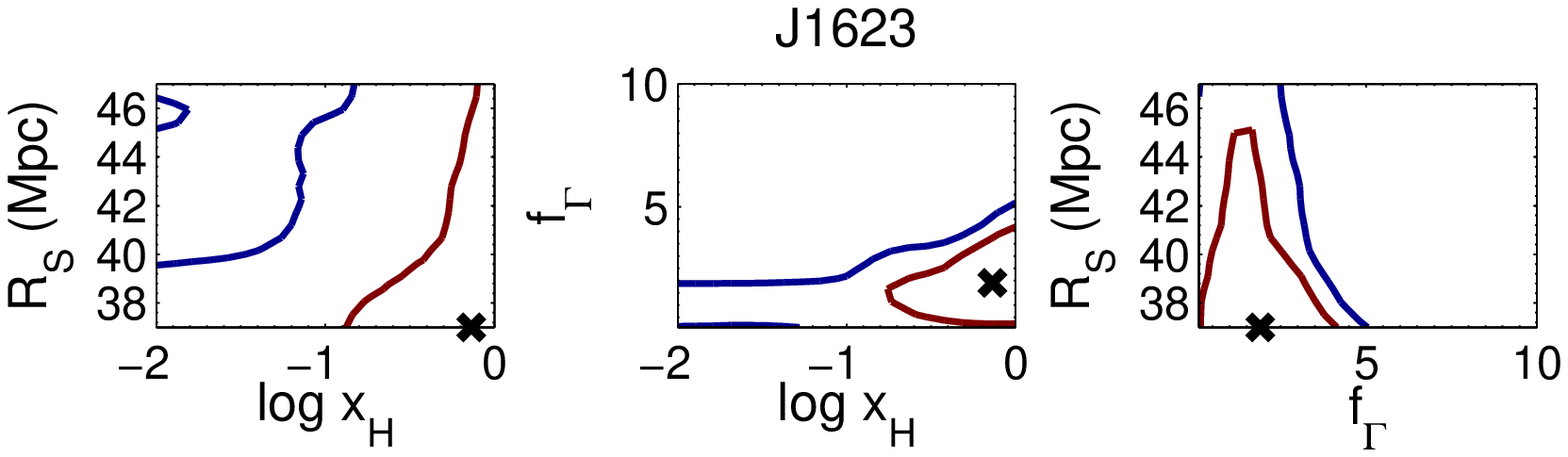}\tabularnewline
\end{tabular}
\caption{\label{fig:uniform} Confidence regions under the assumption
that the IGM is ionized by a spatially uniform background.  The three
rows show the results for the three different quasars. In each row,
the three different panels show the confidence contours in three
different 2D planes, marginalized over the third parameter.  The
contours enclose 68 and 95\% of the marginalized 2D likelihood. The
location of the best-fit model is marked by an ``x'' in each panel.}
\end{figure}

\begin{figure}
\begin{tabular}{c}
\includegraphics[scale=0.75]{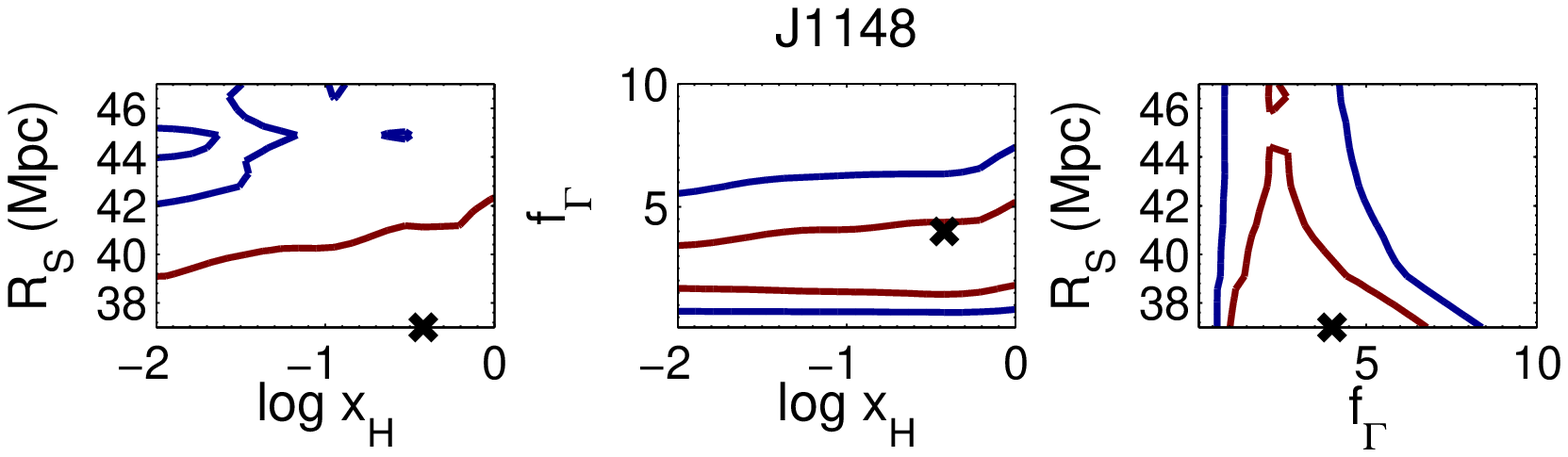}\tabularnewline
\includegraphics[scale=0.75]{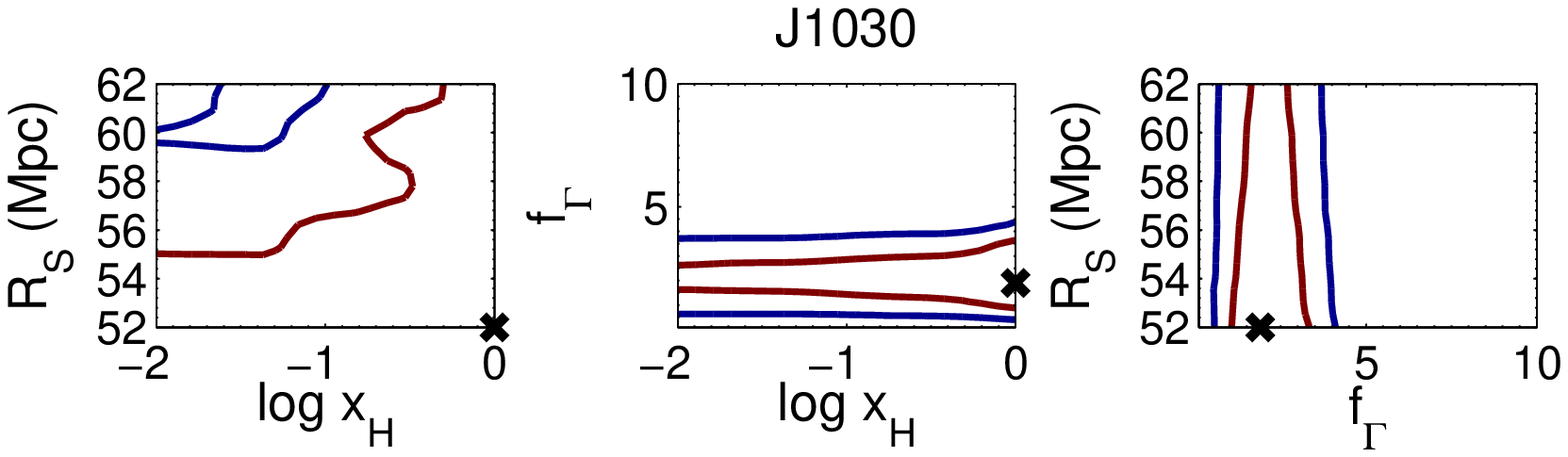}\tabularnewline
\includegraphics[scale=0.75]{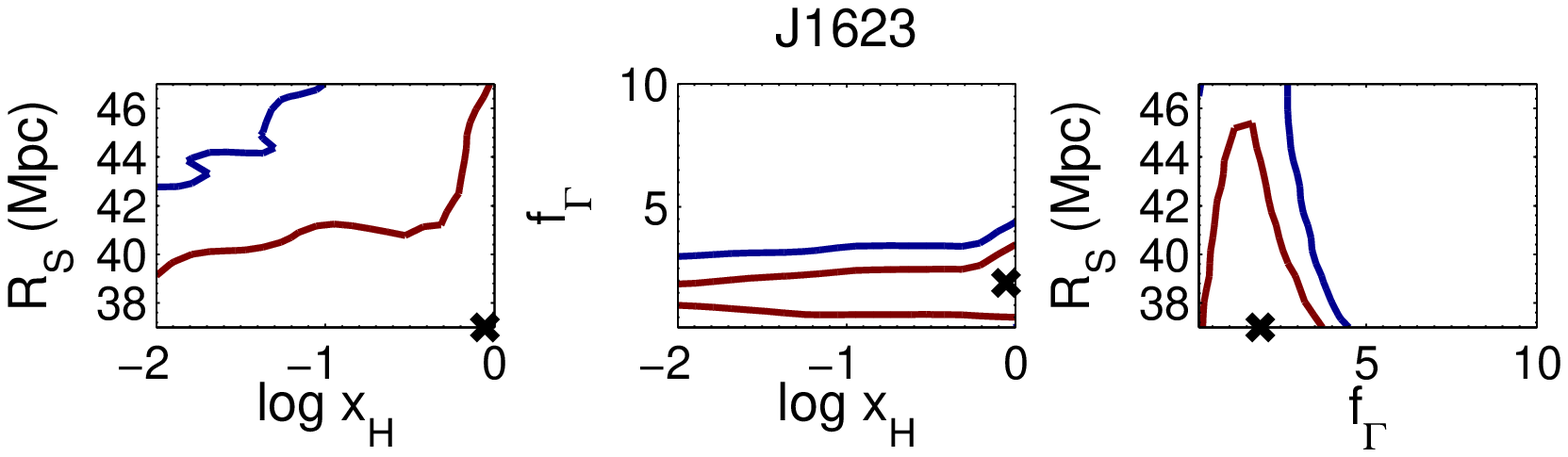}\tabularnewline
\end{tabular}
\caption{\label{fig:patchy} Confidence regions as in
Figure~\ref{fig:uniform}, except in our patchy reionization models.}
\end{figure}

\begin{figure}
\includegraphics[scale=0.75]{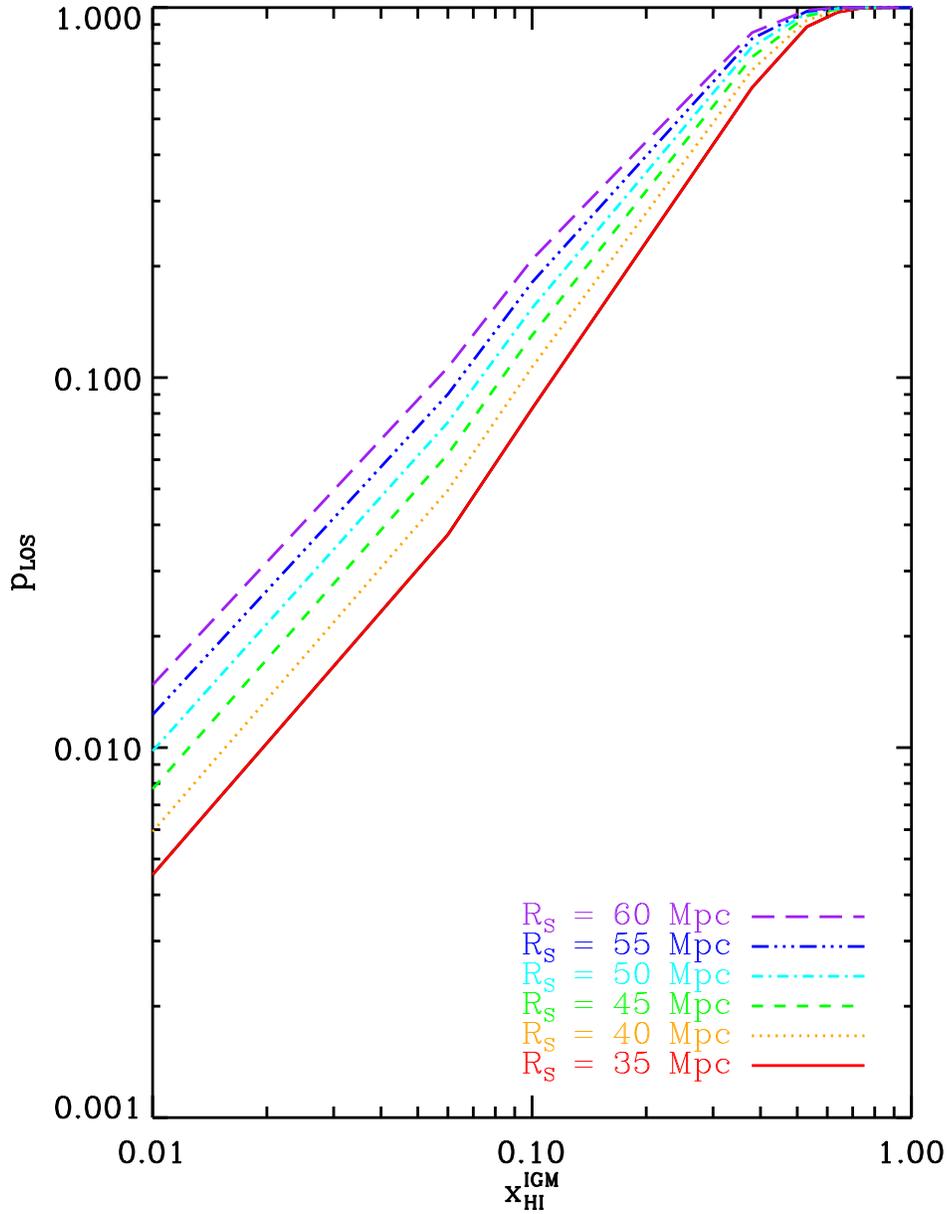}
\caption{\label{fig:Rsprior} Fraction of the simulated lines-of-sight
  ($p_{\rm{LOS}}$) that intersect a neutral pixel within a given
  distance away from the quasar's host halo, as labeled.  These
  fractions are obtained from our patchy reionization models, and
  shown as a function of the global average neutral fraction $\avenf$.
  The figure illustrates that intersecting a neutral patch within
  35-60 (comoving) Mpc from the quasar is much less likely if the IGM
  is highly ionized.}
\end{figure}

\begin{figure}
\begin{tabular}{c}
\includegraphics[scale=0.75]{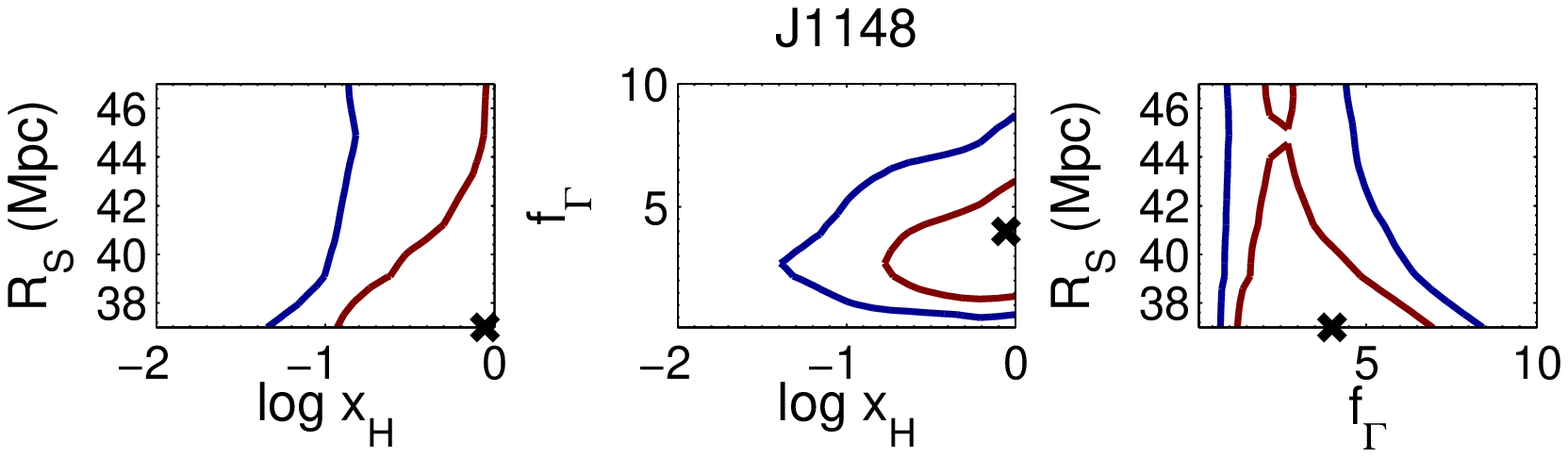}\tabularnewline
\includegraphics[scale=0.75]{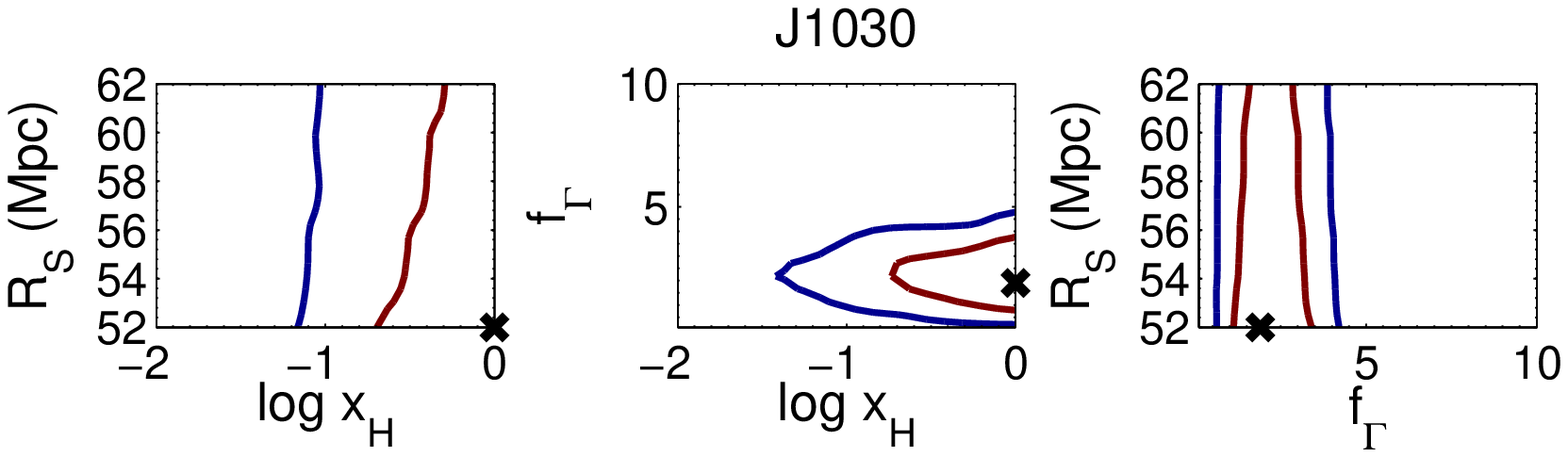}\tabularnewline
\includegraphics[scale=0.75]{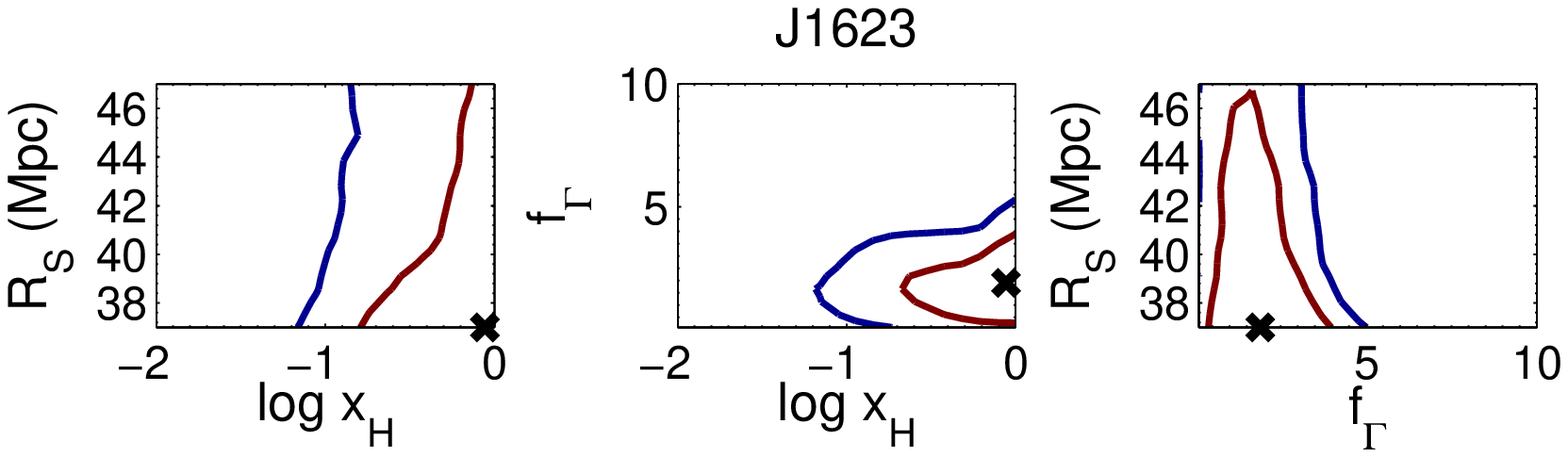}\tabularnewline
\end{tabular}
\caption{\label{fig:patchyprior} Confidence regions as in
Figure~\ref{fig:patchy}, except priors were added for the probability
of finding a neutral patch at $R_S$, taken from
Figure~\ref{fig:Rsprior}.}
\end{figure}

\begin{figure}
\begin{tabular}{c}
\includegraphics[scale=0.75]{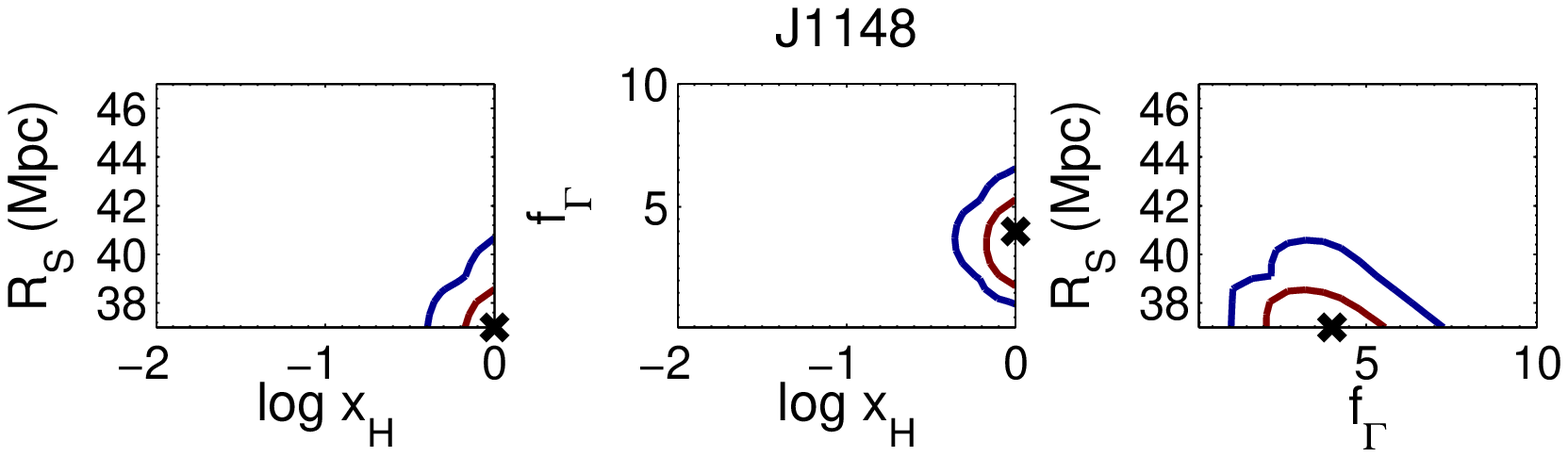}\tabularnewline
\includegraphics[scale=0.75]{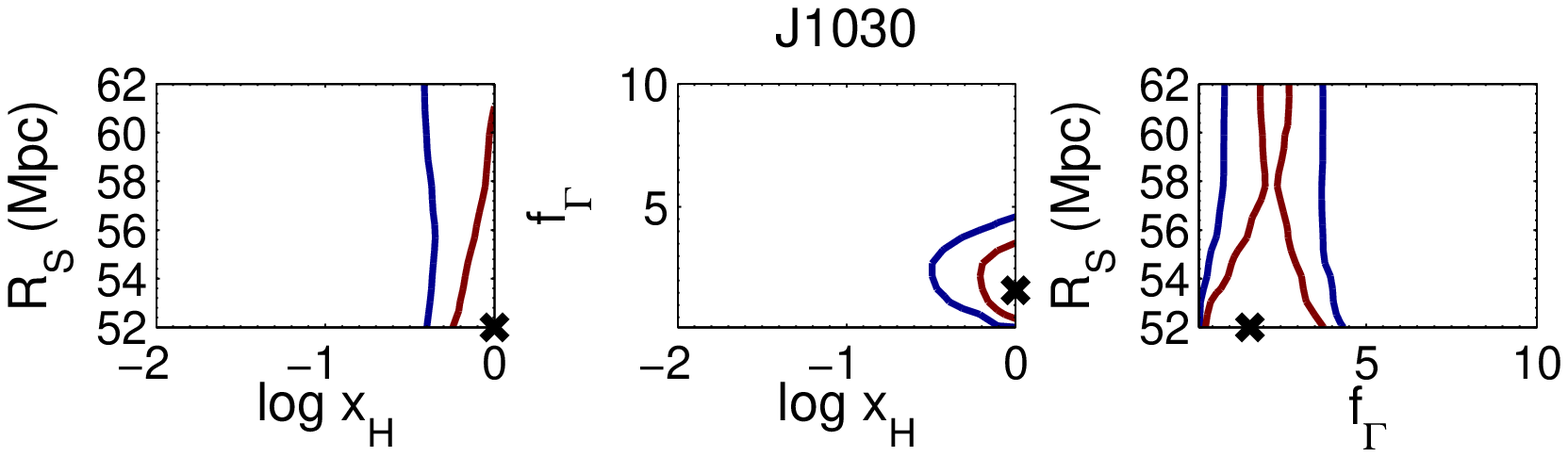}\tabularnewline
\includegraphics[scale=0.75]{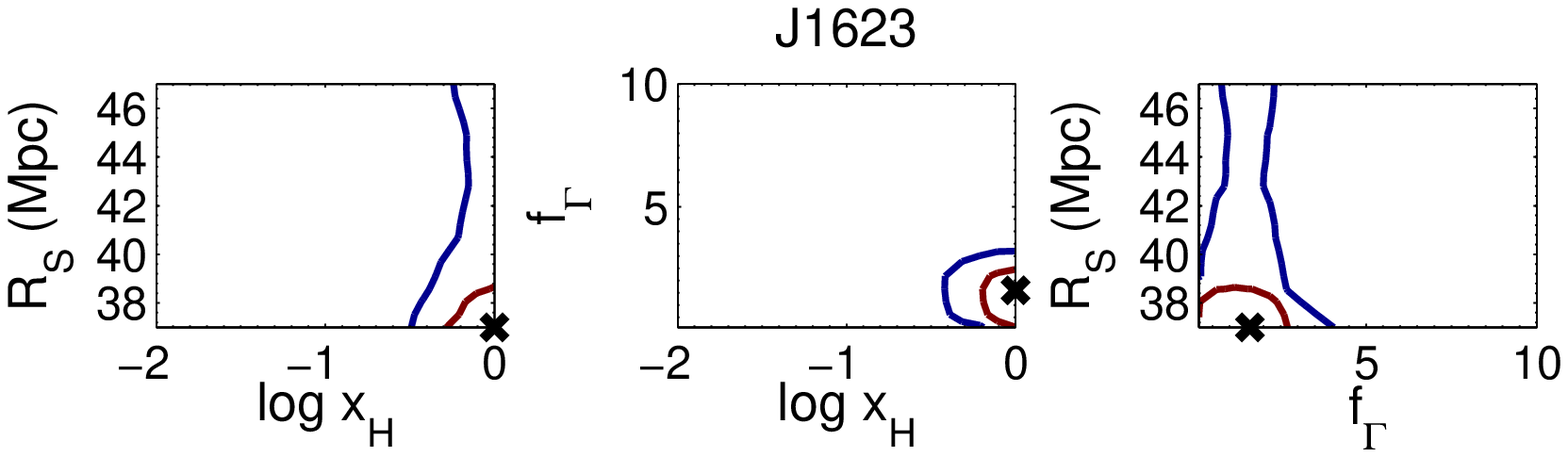}\tabularnewline
\end{tabular}
\caption{\label{fig:patchypriorboot} Confidence regions as in
Figure~\ref{fig:patchyprior}, except that the parametric bootstrapping
procedure, discussed in \S~\ref{subsubsec:parameterfitting}, was used
to estimate the confidence contours.}
\end{figure}

The confidence contours presented in Figures~~\ref{fig:uniform} and
\ref{fig:patchy} illustrate a number of expected basic trends.  Most
importantly for our purposes, as the value of $\avenf$ increases, the
GP damping wing becomes stronger, and, in particular, dominates
further blueward.  In the patchy models, the dependence on $\avenf$
arises from the width of the neutral slab(s) beyond $R>R_S$. This is
different from the uniformly ionized models, in which the damping wing
optical depth simply scales linearly with $\avenf$, at every
frequency.  As $f_{\Gamma}$ increases, the resonant absorption inside
$R<R_S$ is suppressed, and, as $R_S$ increases, the transmission
window generally becomes larger.  Degeneracies are therefore expected:
a larger $R_S$ can be partly compensated by a stronger damping wing
(larger $\avenf$) and/or resonant absorption (a smaller $f_{\Gamma}$).
Likewise, although their wavelength-dependence is different, in
general, a weaker damping wing can be compensated, at least partly, by
stronger resonant absorption.  These trends are indeed visible in
Figures~\ref{fig:uniform} and \ref{fig:patchy} (although tempered by
the priors on $f_{\Gamma}$ and on $R_S$).

\subsection{Damped Lyman Alpha Systems}
\label{sec:res_DLA}

We have thus far considered GP absorption from neutral material in the
IGM (i.e. at densities near the mean background density).  Here we
consider an alternative scenario, in which the IGM is highly ionized,
and the apparently black GP trough in the quasar spectrum is caused
instead by a single discrete absorber, i.e. a DLA, along the LOS.

We consider DLAs with the range of column densities of $\NHI =
\{10^{19}, 5\times10^{19}, 10^{20}, 5\times10^{20}, 10^{21},
5\times10^{21}\}\,\rm{cm}^{-2}$, placed at the same distances $R_S$ as
we used for the edges of the surrounding HII regions in our fiducial
analysis above. We create mock absorption spectra as outlined above,
but replace the IGM damping wings from our reionization simulations
with DLA damping wings.  There are no neutral patches in this model,
and the ionization field is taken to be the sum of the quasar's
ionizing flux, parameterized by $f_\Gamma$ as before, and the flux of
the background galaxies, which is taken from the semi-numerical
simulations as described above. Also as before, the flux is
conservatively assumed to be zero beyond the DLA's location
($R>R_S$). This assumes that the DLA casts a full shadow of the quasar
along the LOS, and makes our constraints independent of the absorption
statistics of the background IGM, away from the quasar.

\begin{figure}
\begin{tabular}{c}
\includegraphics[scale=0.75]{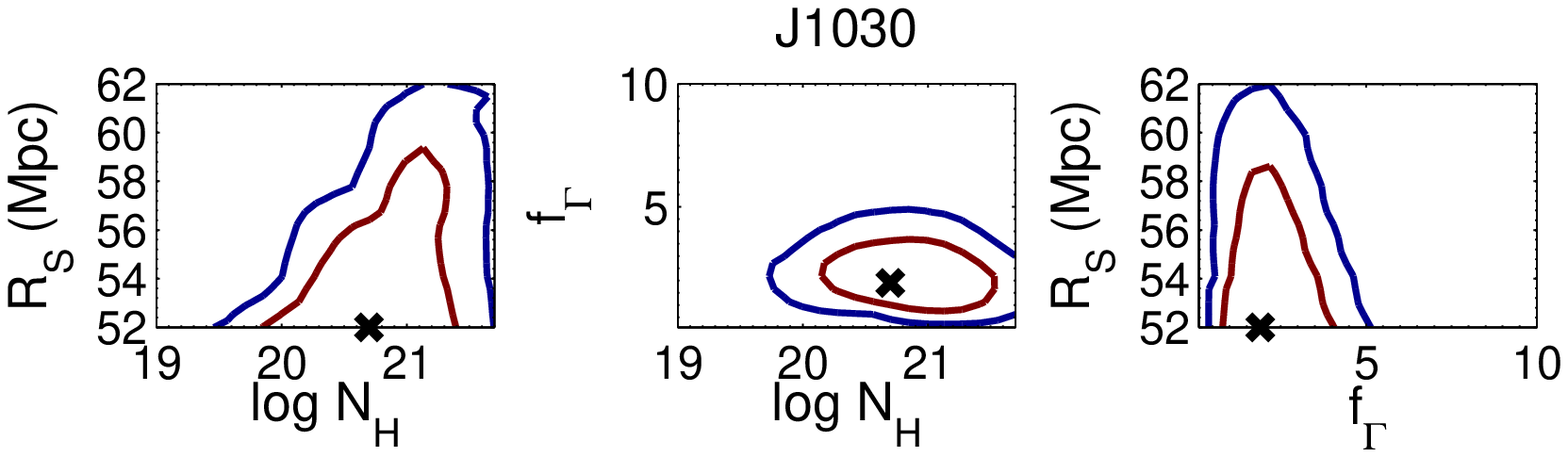}\tabularnewline
\includegraphics[scale=0.75]{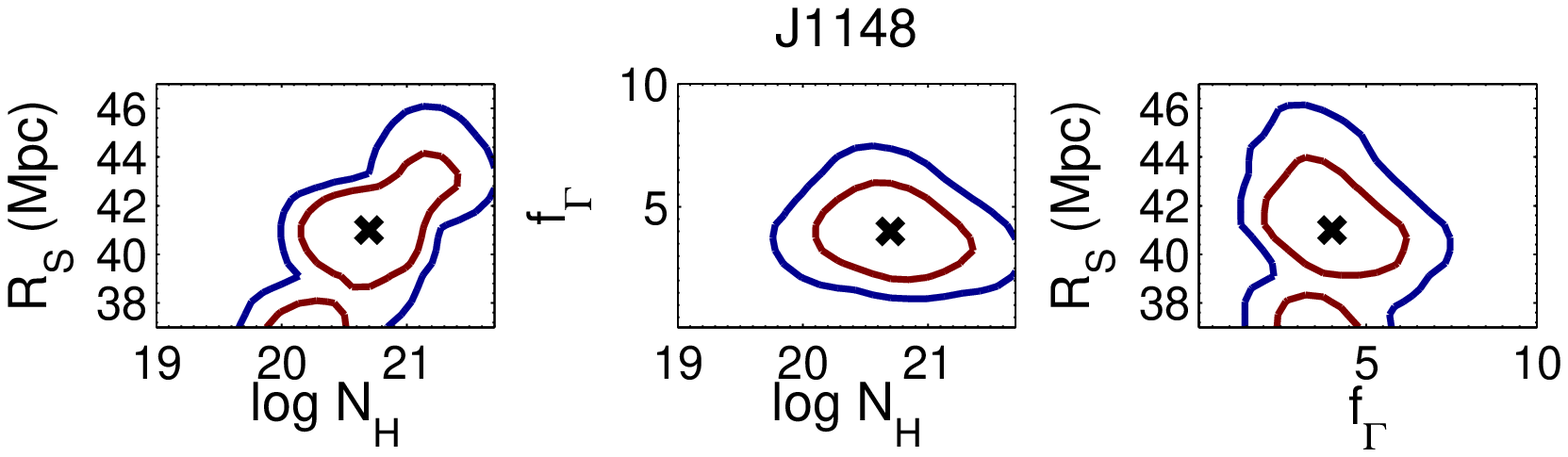}\tabularnewline
\includegraphics[scale=0.75]{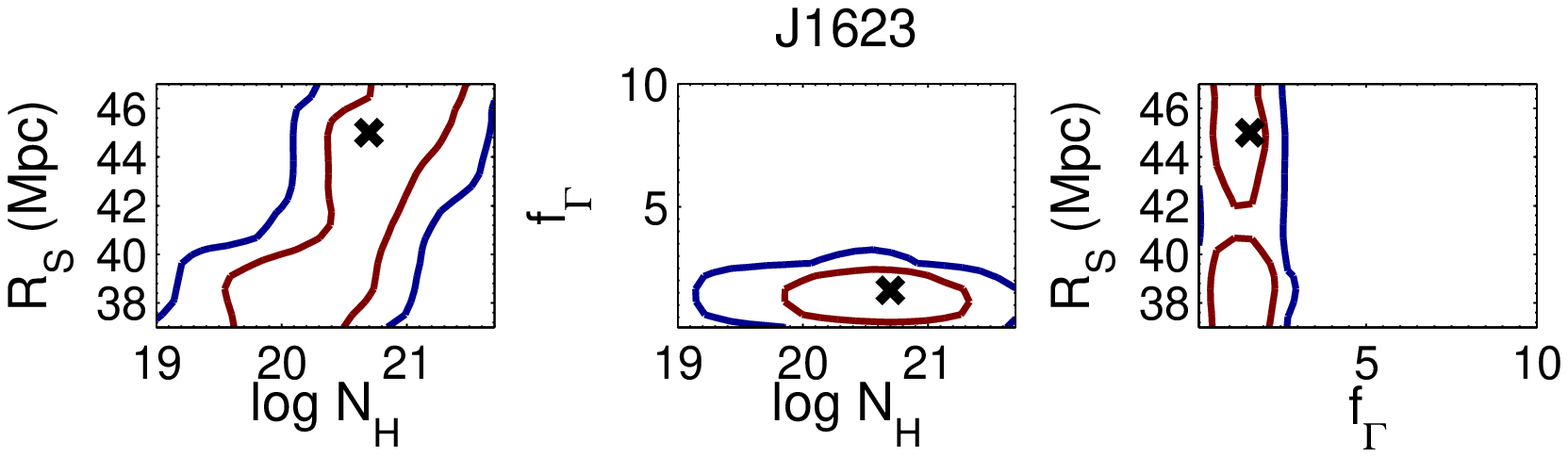}\tabularnewline
\end{tabular}
\caption{\label{fig:dla} Confidence regions in the DLA models.}
\end{figure}

In Figure~\ref{fig:dla}, we present joint two-dimensional confidence
contours in this DLA model.  As the figure shows, all three quasars
require DLAs with very large column densities, with the same best-fit
value of $\NHI = 5\times 10^{20}~{\rm cm^2}$.  As Table~\ref{table1}
further shows, the bootstrapping procedure yields 68\% CL lower limits
on $\NHI$ (marginalized over both $R_S$ and $f_{\Gamma}$) that range
between $\approx 2\times 10^{19} - 3 \times 10^{20}~{\rm cm^2}$.
Interestingly, Table~\ref{table1} also shows that for \qnamestwoeight,
the lower limits on $\NHI$ from the Bayesian and the bootstrapping
analyses are similar, but for \qnamesfourtwo\ and \qnamestwotwo, the
Bayesian limits are much weaker and closer to $\NHI\sim 10^{19}~{\rm
cm^{-2}}$.  This is similar to the differences we found for the lower
limits on $\avenf$. However, unlike in the patchy reionization models,
the DLA damping wings have, by definition, no scatter to account for
this.  To investigate the origin of this difference, we compared a
sub-sample of ``good-fit'' LOSs --- i.e. LOSs having low values of
$D_{KS}<\bar{D}_{KS,obs}$ --- in our low--$\NHI$ model (with
$\NHI=10^{19}~{\rm cm^{-2}}$) to the full sample of $2\times10^4$
sightlines.  The low--$D_{KS}$ sub-sample had noticeably stronger
resonant absorption than the full sample. We thus conclude that the
difference between the Bayesian and bootstrapped lower limits for the
DLA column densities is due to the scatter in the resonant absorption.
This scatter is relatively more important in the low--$\NHI$ models,
and can produce LOSs that mimic stronger DLAs.

Most importantly, these results show that DLAs along the LOS can
provide a plausible explanation of the spectra themselves.  They could
produce the black GP troughs, and, as shown in Table~\ref{table1}, the
probabilities of the best-fit models are comparable to those in the
models with incomplete reionization.  However, finding DLAs with the
high required column densities, relatively near the quasars, is
improbable.  Because of their rarity, the abundance of DLAs at high
redshift is poorly known.  Nevertheless, a modest extrapolation of
existing DLA measurements between $4<z<6$
(\citealt{ProchaskaWolfe09,SongailaCowie10}) to $z\approx 6.5$ imply
the abundance $\sim 0.05$ DLAs with column densities $\NHI \geq
2\times 10^{20}~{\rm cm^2}$ per unit redshift (see, e.g., Fig.~8 and
Eq.~4 in \citealt{SongailaCowie10}).  In our case, $\sim 50$Mpc
corresponds to $\Delta z\sim 0.1$, implying that an abundance of $\sim
10$ DLAs per unit redshift is needed: this is more than two orders of
magnitude higher than the above extrapolation suggests.  In principle,
the abundance of DLAs around quasars could be atypically high;
however, at distances as far as 50 Mpc from the quasar, these biases
should be negligible.  Finally, the high-resolution spectra of these
quasars show no evidence at the required distances of the
low-ionization metal lines typical of DLAs (e.g. \citealt{RPM06,
RPMZ09, Simcoe06}; G. Becker, private communication 2010).

\subsection{The Impact of Ionizing Radiation from Galaxies}
\label{subsec:galaxyflux}

We turn now to the question of whether the ionizing radiation from the
galaxies in the quasar's ionized region is important for our
conclusions.  Generally, we expect that galaxies could be important
for a combination of small $f_\Gamma$ (so that the quasar does not
outshine the galaxies), small $\avenf$ or $\NHI$ (so that the damping
wing absorption, which is independent of the galaxy background, does
not dominate over the resonant absorption), and large $R_S$ (since the
galaxy flux is only weakly dependent on $R$ within $R<R_S$, while the
quasar flux drops as $\sim 1/R^2$)

To investigate the importance of galaxies more explicitly, we have
re-run our analysis, with the ionizing radiation from the galaxies
artificially turned off inside $R<R_S$.  We find that for larger
values of $f_\Gamma$ the galaxy flux can be ignored without much
consequence.  Furthermore, the presence of the galaxy flux has almost
no effect on the best-fit locations or the peak likelihoods, with the
impact restricted to the ``periphery'' of our parameter space, as
expected above.  In general, the contours tend to expand toward larger
$R_S$ and smaller $\avenf$ when the galaxy flux is not modeled.

As an example, we find that DLA models for J1623+3112 are most
affected by the galaxy flux.  The probability density at ($\NHI =
10^{19} \textrm{cm}^{-2}$, $R_S = 47$ Mpc), integrated over
$f_\Gamma$, is $\sim 3$ times higher when the galaxy contribution to
the ionization radiation is neglected. The effect is reversed in
(high--$\NHI$, low--$R_S$) models where, for example, in the same
quasar for $\NHI = 10^{20} \textrm{cm}^{-2}$ and $R_S = 37$ Mpc, the
probability density (again integrated over $f_\Gamma$) is $\sim 10$
times lower without galaxy fluxes.  Nevertheless, the marginalized 1D
constraints remain robust: without galaxies, following the format of
Table~\ref{table1}, the constraints are $(f_\Gamma,R_S,\log \NHI) =
(1.6^{+2.03}_{-0.53},39^{+8}_{-1.34},20.0^{+0.74}_{-0.61})$ with
$\max\left(p\right) = 0.21$.

\subsection{The Impact of Self-Shielded Absorption Systems along the LOS}
\label{sec:rsquared}

In this section, we consider the possibility that attenuation along the LOS is substantial, causing the QSO's intensity to fall off more rapidly with distance than the approximate $\propto r^{-2}$ profile in equation~(\ref{eq:gamma_q}).   As mentioned above, a DLA or strong LLS along the sightline would be rare, and likely detectable in HIRES spectra.  However, several lower column density systems aligning along the LOS could in principle strongly attenuate the flux, thereby mimicking a damping wing signature.

How degenerate are these two signatures?  To answer this question, we perform an additional set of runs, repeating our fiducial analysis above, except {\it without} a damping wing, but instead attenuating the QSO flux by a factor of $\exp(-r/\lambda_Q)$.
  We vary the effective LOS mean free path, $\lambda_Q$, from 60 Mpc down to 5 Mpc.  We find that in {\it all} of these runs, the peak probabilities are reduced from our fiducial models with damping wings (Table~\ref{table1}). Specifically, for J1623+3112 / J1030+0524 / J1148+5251, the best-fit models have $\lambda_G =$ 50 / 40 / 15 Mpc, with $\max\left(p\right)$ lower than the peak probabilities in the fiducial, damping-wing-included results by 76 / 97 / 34 \%.

To further explore the degeneracy between an altered (i.e. steeper) resonant absorption profile and our fiducial damping-wing absorption, we construct an unphysical, ``worst-case'' model.  We again have no damping wings, but instead we adjust the flux, {\it in each pixel}, such that the {\em average} resonant optical depth precisely matches the total (resonant+damping) optical depth in our best-fit models from Table~\ref{table1}.  We thereby allow an entirely arbitrary and contrived profile of the flux along the line-of-sight. (Indeed, in some cases, this profile is unphysical, as it rises with distance away from the quasar.)  Even in this worst-case scenario, we find that the peak probabilities are still lower than in our fiducial cases, by 23\% for J1148+5251, 11\% for J1030+0524,  and 22\% for J1623+3112.  This preference for the presence of a damping wing implies that the spectral fits are (mildly) sensitive to the {\it smoothness} of the damping wing profile (compared with the resonant absorption profile, which has significant pixel-to-pixel variations).

From both of the above tests, we conclude that even if there is unaccounted-for absorption of the QSO flux along the LOS, the observed spectra still prefer an IGM damping wing profile, at the tens of percents level.

\section{Conclusions}
\label{sec:conclusions}

The main result of this paper is the lower limit on the mean
neutral hydrogen fraction,  $\avenf \gsim 0.1$ (at 95\% CL),
inferred from the spectra of three $z>6$ quasars.  Lower limits were
previously obtained for the same three quasars by MH07, using simpler
assumptions, and applying a less sophisticated statistical analysis.
Our new results strengthen these previous limits.  In particular, we
have found that low neutral fractions are ruled out after relaxing the
major assumption in MH07, of a uniform ionizing background. Unlike the
lower limit obtained by MH07, which was driven by the pixel optical
depth statistics in the spectral fitting alone, our new limit is
driven by our additional modeling of the spatial distribution of
neutral patches.  We note that if we conservatively allow the size of
the surrounding HII region to be a fully free parameter, independent
of $\avenf$, our constraints weaken to $\avenf \gsim 0.03$.

Our results arise through a statistical preference for a GP damping
wing absorption component in the spectra.  We confirm that peak
likelihoods drop by a factor of several to more than an order of
magnitude when the spectra are modeled without the damping wing
component, implying the preference for significant neutral hydrogen in
the IGM.  Although such a damping wing could arise from a high-column
density ($\NHI \gsim {\rm{ few}}\times10^{20}{\rm{cm}^{-2}}$) DLA,
such systems are very rare; furthermore, high-resolution spectra show
no evidence, at the required distances, of the low-ionization metal
lines typical of DLAs.  In principle, the imprint of the damping wing could be mimicked by an alternate (not included in our fiducial model) evolution of the resonant absorption with distance from the quasar.
 We
have shown, using a contrived worse-case toy-model, that the
statistical preference for a damping wing can indeed decrease, but
that it can not fully go away, even in this contrived model.  Our
analysis implies that reionization is not yet complete at $z\sim 6.2$.

Finally, we find different results when a Bayesian or a parametric
bootstrapping method is used to estimate the confidence levels on our
model parameters.  We attribute this difference to the fact that the
two methods are sensitive to the sightline-to-sightline scatter in the
absorption statistics in the low-- and the high--$\avenf$ models,
respectively.  The scatter, in particular, in the GP damping wing
strengths and shapes has a strong dependence on the average global
ionization -- as the global neutral fraction increases, the scatter is
diminished. Our results demonstrate that this scatter could provide a
useful additional diagnostic of the ionization topology of the IGM,
when applied simultaneously to a larger quasar sample in the future.

\section{Acknowledgments}

We thank George Becker for discussing the presence of DLAs in HIRES
spectra, and Xiaohui Fan for providing the ESI quasar spectra used in
this paper. We would also like to thank David Spiegel for several
informative discussions.  ZH acknowledges support by the Pol\'anyi
Program of the Hungarian National Office for Research and Technology
(NKTH).

\bibliography{reion}

\begin{thebibliography}{61}
\expandafter\ifx\csname natexlab\endcsname\relax\def\natexlab#1{#1}\fi

\bibitem[{{Barkana}(2002)}]{Barkana02}
{Barkana} R., 2002, New Astronomy, 7, 85

\bibitem[{{Barkana} \& {Loeb}(2004)}]{BLGalaxyFormation2004}
{Barkana} R., {Loeb} A., 2004, \apj, 609, 474

\bibitem[{{Bolton} {et~al}\mbox{.}(2012){Bolton}, {Becker}, {Raskutti},
  {Wyithe}, {Haehnelt}, \& {Sargent}}]{Bolton2012}
{Bolton} J.~S., {Becker} G.~D., {Raskutti} S., {Wyithe} J.~S.~B., {Haehnelt}
  M.~G., {Sargent} W.~L.~W., 2012, \mnras, 419, 2880

\bibitem[{{Bolton} \& {Haehnelt}(2007{\natexlab{a}})}]{BH2007a}
{Bolton} J.~S., {Haehnelt} M.~G., 2007{\natexlab{a}}, \mnras, 374, 493

\bibitem[{{Bolton} \& {Haehnelt}(2007{\natexlab{b}})}]{Bhemiss2007}
---, 2007{\natexlab{b}}, \mnras, 382, 325

\bibitem[{{Bolton} {et~al}\mbox{.}(2011){Bolton}, {Haehnelt}, {Warren},
  {Hewett}, {Mortlock}, {Venemans}, {McMahon}, \& {Simpson}}]{Bolton+11}
{Bolton} J.~S., {Haehnelt} M.~G., {Warren} S.~J., {Hewett} P.~C., {Mortlock}
  D.~J., {Venemans} B.~P., {McMahon} R.~G., {Simpson} C., 2011, \mnras, 416,
  L70

\bibitem[{{Calverley} {et~al}\mbox{.}(2011){Calverley}, {Becker}, {Haehnelt},
  \& {Bolton}}]{CBHB2011}
{Calverley} A.~P., {Becker} G.~D., {Haehnelt} M.~G., {Bolton} J.~S., 2011,
  \mnras, 412, 2543

\bibitem[{{Carilli} {et~al}\mbox{.}(2010){Carilli}, {Wang}, {Fan}, {Walter},
  {Kurk}, {Riechers}, {Wagg}, {Hennawi}, {Jiang}, {Menten}, {Bertoldi},
  {Strauss}, \& {Cox}}]{Carilli+10}
{Carilli} C.~L. {et~al.}, 2010, \apj, 714, 834

\bibitem[{{Cen} \& {Haiman}(2000)}]{CH00}
{Cen} R., {Haiman} Z., 2000, \apjl, 542, L75

\bibitem[{{Crociani} {et~al}\mbox{.}(2011){Crociani}, {Mesinger}, {Moscardini},
  \& {Furlanetto}}]{Crociani11}
{Crociani} D., {Mesinger} A., {Moscardini} L., {Furlanetto} S., 2011, \mnras,
  411, 289

\bibitem[{{Croft}(1998)}]{Croft98}
{Croft} R.~A.~C., 1998, in Eighteenth Texas Symposium on Relativistic
  Astrophysics, {A.~V.~Olinto, J.~A.~Frieman, \& D.~N.~Schramm}, ed., pp.
  664--+

\bibitem[{{Dayal}, {Maselli} \& {Ferrara}(2011){Dayal}, {Maselli}, \&
  {Ferrara}}]{DMF11}
{Dayal} P., {Maselli} A., {Ferrara} A., 2011, \mnras, 410, 830

\bibitem[{{Dijkstra}, {Mesinger} \& {Wyithe}(2011){Dijkstra}, {Mesinger}, \&
  {Wyithe}}]{DMW11}
{Dijkstra} M., {Mesinger} A., {Wyithe} J.~S.~B., 2011, \mnras, 414, 2139

\bibitem[{{Fan}, {Carilli} \& {Keating}(2006){Fan}, {Carilli}, \&
  {Keating}}]{FCK06}
{Fan} X., {Carilli} C.~L., {Keating} B., 2006, \araa, 44, 415

\bibitem[{{Furlanetto}, {Zaldarriaga} \& {Hernquist}(2004){Furlanetto},
  {Zaldarriaga}, \& {Hernquist}}]{FZH2004}
{Furlanetto} S.~R., {Zaldarriaga} M., {Hernquist} L., 2004, \apj, 613, 1

\bibitem[{{Gallerani}, {Choudhury} \& {Ferrara}(2006){Gallerani}, {Choudhury},
  \& {Ferrara}}]{GCF06}
{Gallerani} S., {Choudhury} T.~R., {Ferrara} A., 2006, \mnras, 370, 1401

\bibitem[{{Gunn} \& {Peterson}(1965)}]{GP1965}
{Gunn} J.~E., {Peterson} B.~A., 1965, \apj, 142, 1633

\bibitem[{{Haiman}(2011)}]{Haiman11}
{Haiman} Z., 2011, \nat, 472, 47

\bibitem[{{Komatsu}(2010)}]{WMAP7}
{Komatsu} E. e.~a., 2010, ApJ, submitted, e-print arXiv:1001.4538

\bibitem[{{Kramer} \& {Haiman}(2009)}]{KH09}
{Kramer} R.~H., {Haiman} Z., 2009, \mnras, 400, 1493

\bibitem[{{Lawrence} {et~al}\mbox{.}(2007){Lawrence}, {Warren}, {Almaini},
  {Edge}, {Hambly}, {Jameson}, {Lucas}, {Casali}, {Adamson}, {Dye}, {Emerson},
  {Foucaud}, {Hewett}, {Hirst}, {Hodgkin}, {Irwin}, {Lodieu}, {McMahon},
  {Simpson}, {Smail}, {Mortlock}, \& {Folger}}]{UKIDDS}
{Lawrence} A. {et~al.}, 2007, \mnras, 379, 1599

\bibitem[{{Lidz} {et~al}\mbox{.}(2007{\natexlab{a}}){Lidz}, {McQuinn},
  {Zaldarriaga}, {Hernquist}, \& {Dutta}}]{Lidz+07}
{Lidz} A., {McQuinn} M., {Zaldarriaga} M., {Hernquist} L., {Dutta} S.,
  2007{\natexlab{a}}, \apj, 670, 39

\bibitem[{{Lidz} {et~al}\mbox{.}(2007{\natexlab{b}}){Lidz}, {McQuinn},
  {Zaldarriaga}, {Hernquist}, \& {Dutta}}]{Lidz2007}
---, 2007{\natexlab{b}}, \apj, 670, 39

\bibitem[{{Madau} \& {Rees}(2000)}]{MR00}
{Madau} P., {Rees} M.~J., 2000, \apjl, 542, L69

\bibitem[{{Maselli}, {Ferrara} \& {Gallerani}(2009){Maselli}, {Ferrara}, \&
  {Gallerani}}]{MFG2009}
{Maselli} A., {Ferrara} A., {Gallerani} S., 2009, \mnras, 395, 1925

\bibitem[{{McQuinn} {et~al}\mbox{.}(2007){McQuinn}, {Lidz}, {Zahn}, {Dutta},
  {Hernquist}, \& {Zaldarriaga}}]{McQuinn07}
{McQuinn} M., {Lidz} A., {Zahn} O., {Dutta} S., {Hernquist} L., {Zaldarriaga}
  M., 2007, \mnras, 377, 1043

\bibitem[{{McQuinn} {et~al}\mbox{.}(2008){McQuinn}, {Lidz}, {Zaldarriaga},
  {Hernquist}, \& {Dutta}}]{McQuinn+08}
{McQuinn} M., {Lidz} A., {Zaldarriaga} M., {Hernquist} L., {Dutta} S., 2008,
  \mnras, 388, 1101

\bibitem[{{Mesinger}(2010)}]{M2010}
{Mesinger} A., 2010, \mnras, 407, 1328

\bibitem[{{Mesinger} \& {Dijkstra}(2008)}]{MD2008}
{Mesinger} A., {Dijkstra} M., 2008, \mnras, 390, 1071

\bibitem[{{Mesinger} \& {Furlanetto}(2007)}]{MF2007}
{Mesinger} A., {Furlanetto} S., 2007, \apj, 669, 663

\bibitem[{{Mesinger} \& {Furlanetto}(2009{\natexlab{a}})}]{MF2009}
---, 2009{\natexlab{a}}, \mnras, 400, 1461

\bibitem[{{Mesinger} \& {Furlanetto}(2009{\natexlab{b}})}]{MF09}
---, 2009{\natexlab{b}}, \mnras, 400, 1461

\bibitem[{{Mesinger}, {Furlanetto} \& {Cen}(2011){Mesinger}, {Furlanetto}, \&
  {Cen}}]{21CMFAST}
{Mesinger} A., {Furlanetto} S., {Cen} R., 2011, \mnras, 411, 955

\bibitem[{{Mesinger} \& {Furlanetto}(2008)}]{MF2008b}
{Mesinger} A., {Furlanetto} S.~R., 2008, \mnras, 385, 1348

\bibitem[{{Mesinger} \& {Haiman}(2004)}]{MH2004}
{Mesinger} A., {Haiman} Z., 2004, \apjl, 611, L69

\bibitem[{{Mesinger} \& {Haiman}(2007)}]{MH2007}
---, 2007, \apj, 660, 923

\bibitem[{{Mesinger}, {Haiman} \& {Cen}(2004){Mesinger}, {Haiman}, \&
  {Cen}}]{MHC2004}
{Mesinger} A., {Haiman} Z., {Cen} R., 2004, \apj, 613, 23

\bibitem[{{Mesinger}, {McQuinn} \& {Spergel}(2011){Mesinger}, {McQuinn}, \&
  {Spergel}}]{MMS12}
{Mesinger} A., {McQuinn} M., {Spergel} D., 2011, ArXiv e-prints:1112.1820

\bibitem[{{Miralda-Escud\'e}(1998)}]{ME1998}
{Miralda-Escud\'e} J., 1998, \apj, 501, 15

\bibitem[{{Miralda-Escud{\'e}}(2003)}]{ME2003}
{Miralda-Escud{\'e}} J., 2003, \apj, 597, 66

\bibitem[{{Mortlock} {et~al}\mbox{.}(2011){Mortlock}, {Warren}, {Venemans},
  {Patel}, {Hewett}, {McMahon}, {Simpson}, {Theuns}, {Gonz{\'a}les-Solares},
  {Adamson}, {Dye}, {Hambly}, {Hirst}, {Irwin}, {Kuiper}, {Lawrence}, \&
  {R{\"o}ttgering}}]{Mortlock+11}
{Mortlock} D.~J. {et~al.}, 2011, \nat, 474, 616

\bibitem[{{Osterbrock}(1974)}]{Osterbrock}
{Osterbrock} D.~E., 1974, {Astrophysics of gaseous nebulae}

\bibitem[{{Peebles}(1993)}]{Peebles1993}
{Peebles} P.~J.~E., 1993, {Principles of physical cosmology}. Princeton Series
  in Physics, Princeton, NJ: Princeton University Press, |c1993

\bibitem[{{Phillipps}, {Horleston} \& {White}(2002){Phillipps}, {Horleston}, \&
  {White}}]{PHW2002}
{Phillipps} S., {Horleston} N.~J., {White} A.~C., 2002, \mnras, 336, 587

\bibitem[{{Press}, {Rybicki} \& {Schneider}(1993){Press}, {Rybicki}, \&
  {Schneider}}]{Press1993}
{Press} W.~H., {Rybicki} G.~B., {Schneider} D.~P., 1993, \apj, 414, 64

\bibitem[{{Press} {et~al}\mbox{.}(1992){Press}, {Teukolsky}, {Vetterling}, \&
  {Flannery}}]{NumRec1992}
{Press} W.~H., {Teukolsky} S.~A., {Vetterling} W.~T., {Flannery} B.~P., 1992,
  {Numerical recipes in FORTRAN. The art of scientific computing}. Cambridge:
  University Press, |c1992, 2nd ed.

\bibitem[{{Prochaska} \& {Wolfe}(2009)}]{ProchaskaWolfe09}
{Prochaska} J.~X., {Wolfe} A.~M., 2009, \apj, 696, 1543

\bibitem[{{Ryan-Weber}, {Pettini} \& {Madau}(2006){Ryan-Weber}, {Pettini}, \&
  {Madau}}]{RPM06}
{Ryan-Weber} E.~V., {Pettini} M., {Madau} P., 2006, \mnras, 371, L78

\bibitem[{{Ryan-Weber} {et~al}\mbox{.}(2009){Ryan-Weber}, {Pettini}, {Madau},
  \& {Zych}}]{RPMZ09}
{Ryan-Weber} E.~V., {Pettini} M., {Madau} P., {Zych} B.~J., 2009, \mnras, 395,
  1476

\bibitem[{{Simcoe}(2006)}]{Simcoe06}
{Simcoe} R.~A., 2006, \apj, 653, 977

\bibitem[{{Songaila} \& {Cowie}(2002)}]{SC02}
{Songaila} A., {Cowie} L.~L., 2002, \aj, 123, 2183

\bibitem[{{Songaila} \& {Cowie}(2010)}]{SongailaCowie10}
---, 2010, \apj, 721, 1448

\bibitem[{{Telfer} {et~al}\mbox{.}(2002){Telfer}, {Zheng}, {Kriss}, \&
  {Davidsen}}]{TZKD2002}
{Telfer} R.~C., {Zheng} W., {Kriss} G.~A., {Davidsen} A.~F., 2002, \apj, 565,
  773

\bibitem[{{Trac} \& {Cen}(2007)}]{TC07}
{Trac} H., {Cen} R., 2007, \apj, 671, 1

\bibitem[{{White} {et~al}\mbox{.}(2003){White}, {Becker}, {Fan}, \&
  {Strauss}}]{WBFS2003}
{White} R.~L., {Becker} R.~H., {Fan} X., {Strauss} M.~A., 2003, \aj, 126, 1

\bibitem[{{Willott} {et~al}\mbox{.}(2010{\natexlab{a}}){Willott}, {Albert},
  {Arzoumanian}, {Bergeron}, {Crampton}, {Delorme}, {Hutchings}, {Omont},
  {Reyl{\'e}}, \& {Schade}}]{Willott+10b}
{Willott} C.~J. {et~al.}, 2010{\natexlab{a}}, \aj, 140, 546

\bibitem[{{Willott} {et~al}\mbox{.}(2010{\natexlab{b}}){Willott}, {Delorme},
  {Reyl{\'e}}, {Albert}, {Bergeron}, {Crampton}, {Delfosse}, {Forveille},
  {Hutchings}, {McLure}, {Omont}, \& {Schade}}]{Willott+10}
---, 2010{\natexlab{b}}, \aj, 139, 906

\bibitem[{{Wyithe} \& {Loeb}(2007)}]{WL2007}
{Wyithe} J.~S.~B., {Loeb} A., 2007, \mnras, 374, 960

\bibitem[{{Zahn} {et~al}\mbox{.}(2011{\natexlab{a}}){Zahn} {et~al.}}]{Zahn12}
{Zahn} ., {et~al.}, 2011{\natexlab{a}}, ArXiv e-prints:1111.6386

\bibitem[{{Zahn} {et~al}\mbox{.}(2011{\natexlab{b}}){Zahn}, {Mesinger},
  {McQuinn}, {Trac}, {Cen}, \& {Hernquist}}]{Zahn+2011}
{Zahn} O., {Mesinger} A., {McQuinn} M., {Trac} H., {Cen} R., {Hernquist} L.~E.,
  2011{\natexlab{b}}, \mnras, 414, 727

\bibitem[{{Zel'dovich}(1970)}]{Z1970}
{Zel'dovich} Y.~B., 1970, \aap, 5, 84

\end{thebibliography}

\clearpage

\begin{table*}
\begin{tabular}{|c|cc|cc|}
\hline 
Model & $P(D_{KS})$ best-fit & $\max\left(p\right)$ & Bootstrap best-fit & $\max\left(N_{bf}\right)$\\[+0.1in]
\hline
\multicolumn{1}{|c}{} & \multicolumn{4}{c|}{J1148+5251}\\
\hline
uniform & $(4.0^{+2.35}_{-2.86},37^{+7.89}_{-0},0.88^{+0.12}_{-0.84})$ & 0.201 & $(4.0^{+1.31}_{-2.34},37^{+1.05}_{-0},1.00^{+0}_{-0.38})$ & 7595\\[+0.1in]
MH07 & $(1.6,40,0.16)$ & 0.03 & - & - \\[+0.1in]
patchy & $(4.0^{+1.83}_{-2.86},37^{+7.37}_{-0},0.38^{+0.62}_{-0.36})$ & 0.248 & $(4.0^{+1.83}_{-1.82},37^{+7.37}_{-0},0.64^{+0.36}_{-0.496})$ & 4691\\[+0.1in]
$R_S$ prior & $(4.0^{+1.83}_{-2.86},37^{+9.47}_{-0},0.88^{+0.12}_{-0.73})$ & 0.196 & $(4.0^{+1.31}_{-2.34},37^{+2.11}_{-0},1.00^{+0}_{-0.38})$ & 10126\\[+0.1in]
DLA$^*$ & $(4.0^{+2.35}_{-2.86},41^{+6}_{-4},20.7^{+0.72}_{-1.41})$ & 0.149 & $(4.0^{+1.83}_{-1.82},41^{+2.84}_{-4},20.7^{+0.29}_{-0.28})$ & 8332\\[+0.1in]
\hline
\multicolumn{1}{|c}{} & \multicolumn{4}{c|}{J1030+0524}\\
\hline
uniform & $(1.9^{+1.85}_{-0.76},52^{+10}_{-0},0.88^{+0.12}_{-0.81})$ & 0.194 & $(1.6^{+1.63}_{-0.98},52^{+5.79}_{-0},1.00^{+0}_{-0.62})$ & 4293\\[+0.1in]
MH07 & $(1.0,41,1.0)$ & 0.34 & - & - \\[+0.1in]
patchy & $(1.9^{+1.85}_{-0.76},52^{+8.95}_{-0},1.00^{+0}_{-0.984})$ & 0.447 & $(1.6^{+1.63}_{-0.98},52^{+8.42}_{-0},1.00^{+0}_{-0.89})$ & 3466\\[+0.1in]
$R_S$ prior & $(1.9^{+1.85}_{-0.76},52^{+8.95}_{-0},1.00^{+0}_{-0.89})$ & 0.447 & $(1.6^{+1.63}_{-0.98},52^{+10}_{-0},1.00^{+0}_{-0.52})$ & 3363\\[+0.1in]
DLA$^*$ & $(1.9^{+2.89}_{-0.76},54^{+6.95}_{-2},20.7^{+0.86}_{-1.27})$ & 0.259 & $(1.9^{+1.85}_{-1.28},52^{+7.37}_{-0},20.7^{+0.57}_{-1.36})$ & 2563\\[+0.1in]
\hline
\multicolumn{1}{|c}{} & \multicolumn{4}{c|}{J1623+3112}\\
\hline
uniform & $(1.9^{+1.85}_{-1.28},37^{+8.42}_{-0},0.73^{+0.27}_{-0.69})$ & 0.406 & $(1.6^{+0.58}_{-1.5},37^{+2.11}_{-0},1.00^{+0}_{-0.82})$ & 1667\\[+0.1in]
MH07 & $(0.7,29,1.0)$ & 0.39 & - & - \\[+0.1in]
patchy & $(1.9^{+1.33}_{-1.8},37^{+8.95}_{-0},0.88^{+0.12}_{-0.86})$ & 0.472 & $(1.3^{+0.88}_{-0.68},37^{+6.84}_{-0},1.00^{+0}_{-0.93})$ & 1565\\[+0.1in]
$R_S$ prior & $(1.9^{+1.33}_{-1.28},37^{+8.95}_{-0},0.88^{+0.12}_{-0.74})$ & 0.464 & $(1.6^{+0.58}_{-1.5},37^{+10}_{-0},1.00^{+0}_{-0.38})$ & 3290\\[+0.1in]
DLA$^*$ & $(1.9^{+1.85}_{-1.8},41^{+6}_{-4},21^{+0.42}_{-1.86})$ & 0.298 & $(1.6^{+0.58}_{-1.5},45^{+2}_{-8},20.7^{+0.29}_{-0.56})$ & 1166\\[+0.1in]
\hline
\end{tabular}
\caption{\label{table1} The table shows the best-fit parameters in
  different models for the three quasar spectra (along with their 95\%
  errors when appropriate, marginalized over the other two parameters
  in each case). For each quasar, the five rows describe the following
  models: (i) uniform reionization (\S~\ref{sec:res_uniform}), (ii)
  the earlier results of MH07, for reference, (iii) patchy
  reionization (\S~\ref{sec:nopriors}), (iv) patchy reionization with
  a physical prior added for the distance $R_S$ to the nearest neutral
  patch (\S~\ref{sec:priors}), and (v) a model in which the GP trough
  is caused by a DLA (\S~\ref{sec:res_DLA}).  For each model, the four
  rows show (i) the best-fit parameters that maximize the probability
  based on the KS test (the values listed are for
  $(f_{\Gamma},R_S/\rm{ Mpc},\avenf)$ in the reionization models, and
  $(f_{\Gamma},R_S/\rm{ Mpc}, \log\NHI/\rm{ cm}^{-2})$ in the DLA
  models), (ii) the value of this peak probability $p$; (iii) the
  best-fit parameters from the bootstrapping procedure (except MH07
  did not perform any bootstrapping), and (iv) the peak number
  ($N_{bf}$) of bootstrap best--fits out of $2\times10^4$ trials in
  this model.  $^{*}$Note that we report 68\% confidence intervals for
  $\log\NHI$.}
\end{table*}

\clearpage

\clearpage

\clearpage

\clearpage

\clearpage

\clearpage

\clearpage

\end{document}